\documentclass[fleqn,usenatbib]{mnras_tex_edited}

\usepackage{amssymb,amsmath,verbatim,mathtools,needspace,enumitem,etoolbox,graphicx,physics,microtype,natbib,url,hyperref,mathrsfs,bm,ulem}

\renewcommand{\emph}[1]{\textit{#1}}

\newcommand{\approptoinn}[2]{\mathrel{\vcenter{
  \offinterlineskip\halign{\hfil$##$\cr
    #1\propto\cr\noalign{\kern2pt}#1\sim\cr\noalign{\kern-2pt}}}}}

\def\aj{\rm{AJ}}
\def\araa{\rm{ARA\&A}}
\def\apj{\rm{ApJ}}
\def\apjl{\rm{ApJ}}
\def\apjs{\rm{ApJS}}
\def\aap{\rm{A\&A}}

\def\mnras{\rm{MNRAS}}

\def\pasa{\rm{PASA}}

\newcommand{\bham}{School of Physics and Astronomy \&	 Institute for Gravitational Wave Astronomy, University of Birmingham,\vspace{-0.05cm}\\$\;$Birmingham, B15 2TT, UK}
\newcommand{\aei}{Max Planck Institute for Gravitational Physics (Albert Einstein Institute), Am M\"uhlenberg 1, Potsdam 14476, Germany}
\newcommand{\leicester}{School of Physics and Astronomy, University of Leicester, University Road, Leicester LE1 7RH, UK}
\newcommand{\exoplanets}{Centre for Exoplanets and Habitability, University of Warwick, Coventry CV4 7AL, UK}
\newcommand{\warwick}{Department of Physics, University of Warwick, Coventry CV4 7AL, UK}
\newcommand{\milan}{Dipartimento di Fisica ``G. Occhialini'', Universit\'a degli Studi di Milano-Bicocca, Piazza della Scienza 3, 20126 Milano, Italy}
\newcommand{\leiden}{Leiden Observatory, Leiden University, P.O. Box 9513, NL-2300 RA Leiden, The Netherlands}
\newcommand{\infn}{INFN, Sezione di Milano-Bicocca, Piazza della Scienza 3, 20126 Milano, Italy}
\newcommand{\ens}{Univ Lyon, Univ Lyon1, Ens de Lyon, CNRS, Centre de Recherche Astrophysique de Lyon UMR5574,\\ F-69230, Saint-Genis-Laval, France}

\title[Bardeen-Petterson effect in SMBH binaries]{The Bardeen-Petterson effect in accreting supermassive black-hole binaries: disc breaking and critical obliquity}

\author[Nealon et al.]{Rebecca Nealon$^{1,2}$, Enrico Ragusa$^{3,4}$, Davide Gerosa$^{5,6,7}$, Giovanni Rosotti$^{3,8}$, \newauthor Riccardo Barbieri$^9$
\medskip
\\
$^{1}$\exoplanets\\
$^{2}$\warwick\\
$^{3}$\leicester\\
$^{4}$\ens\\
$^{5}$\milan\\
$^{6}$\infn\\
$^{7}$\bham\\
$^{8}$\leiden\\
$^{9}$\aei\\
}

\pubyear{2021}

\begin{document}
\label{firstpage}
\pagerange{\pageref{firstpage}--\pageref{lastpage}}
\maketitle

\begin{abstract}
The inspiral of supermassive black-hole binaries in gas-rich environment is driven by the presence of an accretion disc and viscous interactions tend to align the spin of the black holes with the orbital angular momentum of the disc. Recent work introduced a new  iterative approach to describe the alignment process and the resulting non-linear evolution of the surrounding warped accretion disc. Their model predicted that black-hole spins reach either full alignment or a `critical obliquity' where solutions to the warp equations cease to exist. In this paper, we show that this critical region corresponds to the disc breaking phenomenon, where the disc is disrupted into two or more discrete sections. We use 3D hydrodynamical simulations to (i) recover the predictions of the semi-analytic model and (ii) unveil a richer phenomenology where the disc exhibits either unsuccessful, single and multiple breaks. We additionally identify hydrodynamic effects such as spiral arms that are able to stabilise the disc against breaking beyond criticality. Our results show that when disc breaking occurs, the ability of  black holes and disc to align is compromised and in some cases even prevented as the binary inspirals.
\end{abstract}

\begin{keywords}
accretion, accretion discs --- black-hole mergers --- gravitational waves ---  hydrodynamics
\end{keywords}


\section{Introduction}

Accretion discs play a pivotal role in a variety of astrophysical processes, ranging from planet formation to interacting binary stars, and active-galactic nuclei (AGN) \citep{1981ARA&A..19..137P,2002apa..book.....F}. While accretion onto a single Newtonian object results into a planar disc configuration, the presence of external torques might induce a distorted, or `warped', structure. Known processes that can excite disc warps include higher-order harmonics of the central  gravitational potential \citep{2009AJ....137.3706T}, the presence of a binary companions \citep{2000ApJ...538..326L}, embedding in a stellar clusters \citep{2012ApJ...748...63B}, radiation pressure from the central object \citep{1996MNRAS.281..357P}, magnetic fields \citep{1999ApJ...524.1030L}, as well as relativistic effects \citep{1975ApJ...195L..65B}.  

General-relativistic frame dragging is at the heart of the so-called `Bardeen-Petterson effect'. For a disc surrounding a spinning black hole (BH), Lense-Thirring precession preferentially dissipates angular momentum in a direction perpendicular to the BH spin, thus acting towards aligning the disc with the equatorial plane of the BH \citep{1918PhyZ...19..156L}. Crucially, the relevant precession frequency decreases rather steeply with the distance from the BH ($\Omega_{\rm LT}\propto1/R^3$, where $R$ a radial coordinate; e.g. \citealt{1985MNRAS.213..435K}), implying that momentum can be efficiently re-distributed only for gas rings that are sufficiently close to the BH. While the inner disc ---up to the so-called `warp radius'--- aligns with the BH spin, the outer disc maintains its generically misaligned orientation. However, most of the angular momentum resides in this outer disc that thus reacts by pulling the BH spin toward a fully planar configuration \citep{1976IAUS...73..225R}. For typical AGN-disc parameters, this process takes place on a timescale of $1-10$ Myr \citep{1998ApJ...506L..97N,2013MNRAS.429L..30L}.

The Bardeen-Petterson effect has been invoked to explain misaligned jets in both AGNs \citep{2006ApJ...638..120C,2007MNRAS.379..135C,2010ApJ...713L..74F} and microquasars \citep{2002MNRAS.336.1371M,2008MNRAS.391L..15M}, as well as quasi-periodic oscillations in X-ray binaries \citep{2001ApJ...553..955F} and the light curves of some tidal disruption events \citep{2013ApJ...762...98L}. For discs surrounding supermassive BH binaries, gas-driven spin alignment is thought to be a key process to prevent the ejections of BH-merger remnants from their host galaxies following relativistic recoils \citep{2007ApJ...661L.147B,2013ApJ...774...43M, 2010MNRAS.402..682D,2015MNRAS.451.3941G}. The upcoming gravitational-wave mission LISA \citep{2017arXiv170200786A} has the potential of directly measuring the spin directions of several of these systems, thus providing a complementary probe to further test the occurrence of the Bardeen-Petterson effect~\citep{2008ApJ...684..822B,2014ApJ...794..104S,2021MNRAS.501.2531S}.

Building on earlier explorations by \cite{1996MNRAS.282..291S}, \cite{2007MNRAS.381.1617M,2009MNRAS.400..383M}, and \cite{2014MNRAS.441.1408T}, some of the authors recently presented a systematic investigation of the Bardeen-Petterson effect in accreting supermassive BH binaries  \citep{2020MNRAS.496.3060G}. They put forward a  one-dimensional (1D) numerical scheme that  takes into account, in a consistent fashion, both (i) the non-linear character of the fluid viscosities in warped configurations~\citep{1999MNRAS.304..557O,2013MNRAS.433.2403O} and (ii) the combined effect of the Lense-Thirring and companion torques. Their study highlighted the occurrence of a `critical obliquity' ---a specific region in the parameter space where solutions to the underlying 1D boundary-value problem cease to exist. Hints of this behaviour were previously reported by \cite{2014MNRAS.441.1408T} with a different numerical scheme. \cite{2020MNRAS.496.3060G} conjectured that their numerical divergences correspond to a physical scenario where the disk breaks into disconnected regions, hindering the subsequent spin-alignment process. 

In the context of BH accretion, disc breaking and tearing has almost exclusively been explored using numerical simulations. \citet{2000MNRAS.315..570N} conducted the first three-dimensional (3D) simulations of a misaligned accretion disc around a BH and found that the inner disc aligns with the BH spin and the outer disc maintains its original misalignment, in broad agreement with the theoretical expectations of the Bardeen-Petterson effect. Additionally, for large initial misalignments, they found that the transition between the inner and outer disc plane was no longer continuous so that the disc was `close to breaking into two discrete pieces'  \citep{2000MNRAS.315..570N}. The concept of disc breaking was later expanded upon by \citet{2012ApJ...757L..24N}, who demonstrated that discs could tear into more than just two pieces, and under certain circumstances could break into precessing rings of gas. Breaking was further confirmed in the wave-like regime by \cite{2015MNRAS.448.1526N}, suggesting that disc breaking is an inevitable consequence of moderate to strongly misaligned flows accreting onto rotating BHs. Disc breaking has also been confirmed using both a grid based magneto-hydrodynamic treatment \citep{2021MNRAS.507..983L} and analytic arguments for discs subjected to a non-Keplerian potentials \citep{2020MNRAS.495.1148D}.

In this paper, we investigate the interplay between disc criticality  and disc breaking. Guided by the 1D predictions of \cite{2020MNRAS.496.3060G}, we present a large suite of 3D smoothed particle hydrodynamics (SPH) simulations of misaligned accretion discs surrounding spinning BHs in binary systems. We confirm that (i) the occurrence of a critical obliquity corresponds to disc breaking and (ii) its importance increases as the influence of the BH companion increases. Our simulations further allow us to unveil a richer phenomenology which includes (iii) disc breaking into both single and multiple precessing rings as well as (iv) the stabilizing effect of spiral arms in the disc.  A future publication will make use of these prescriptions to investigate the spin directions of large populations of supermassive BHs and their relevance to the LISA mission.

This paper is organised as follows. In Sec.~\ref{secdiscbreaking} we briefly summarise the physics of warped accretion discs and the breaking conditions we employ. In Sec.~\ref{secnumerics}, we present our numerical implementation. In Sec.~\ref{secresults}, we illustrate our main results in terms of both disc morphology and BH spin alignment. In Sec.~\ref{section:discussion}~and~\ref{secconclusions} we present our conclusions and highlight prospects for future work in this area.

\section{Disc breaking}
\label{secdiscbreaking}

In this section we summarise the relevant analytic framework of warped discs around BHs that we will make use of in this work. Consistent with most of the previous literature (but see \citealt{2021ApJ...909...82R}), we use the terms `breaking' and `tearing' interchangeably to both describe discs that separate into a discontinuous structure.

\subsection{Defining a break}
\label{section:defining_a_break}
To define where and when the disc tears in our simulations we consider both the mass surface density and warp profiles. First, we require that the surface density profile $\Sigma(R)$ present a sustained local minimum $\Sigma_{\rm min}$. This straightforward definition was also used by both    \citet{2012ApJ...757L..24N} and \citet{2015MNRAS.448.1526N}. 
Second, we consider the gradient of the angular momentum profile (also refereed to as `warp profile', e.g.~\citealt{2010MNRAS.405.1212L})
\begin{align}
    \psi(R) = R \abs{ \frac{\partial \hat{\mathbfit{L}}(R)}{\partial R}},
    \label{equation:psi}
\end{align}
where $\hat{\mathbfit{L}}$ is the unit vector pointing in the direction of the disc angular momentum. We require an increasing local maximum $\psi_{\rm max}$ to determine that disc breaking is occurring. This is similar to previous work by both \citet{2000MNRAS.315..570N}, who considered a steep increase in the radial warp profile to identify a potential break, and \cite{2021ApJ...909...82R}.

\begin{figure*}
    \centering
    \includegraphics[width=\textwidth]{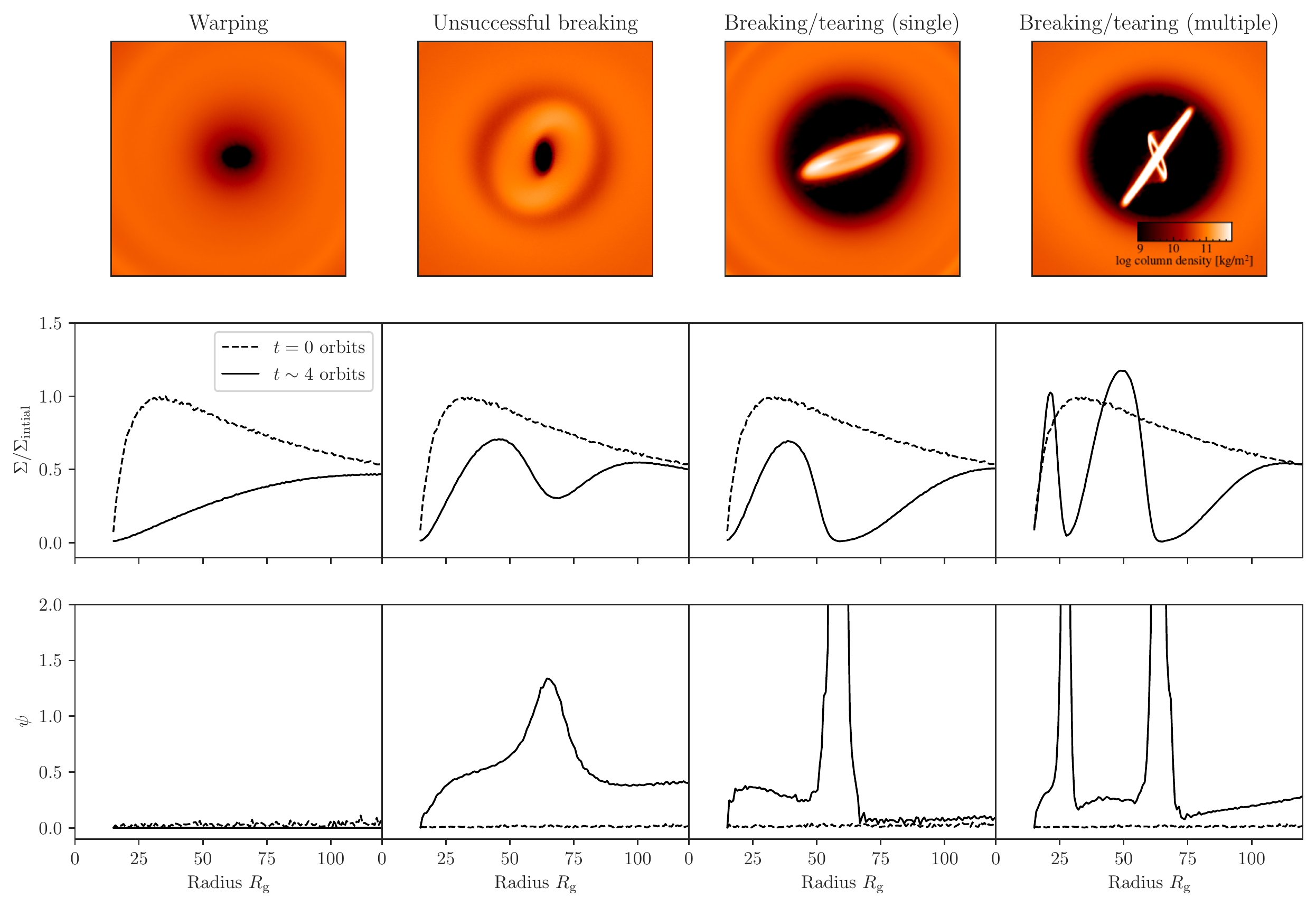}
    \caption{The different disc structure in a simulations with warping, unsuccessful breaking, successful breaking/tearing with one ring and successful breaking/tearing with multiple rings (left to right). The renderings in the top row show the $x$-$y$ plane, with the BH oriented such that $\bm{J}=(\sin \theta, 0, \cos \theta)$. The second and third rows show the surface density $\Sigma$ and  the warp profile  $\psi$. The simulations used here, from left to right, have parameters $\kappa=3.9\times 10^{-5}$ and $\theta = 60\degr$, $\kappa=1.2$ and $\theta = 40\degr$, $\kappa=1.4\times 10^{-4}$ and $\theta = 80\degr$,  $\kappa=2.6\times 10^{-1}$ and $\theta = 60\degr$.
     All simulations are shown at $t=4.3$ companion orbits. }
    \label{fig:tearing_example}
\end{figure*}

With these two metrics in mind, we find that the outcome of our simulations can be classified into the following outcomes. These four scenarios are showcased in Fig.~\ref{fig:tearing_example} and a description of how we discriminate these can be found in Sec.~\ref{section:diditbreak}.
\begin{enumerate}[leftmargin=*]
    \item \emph{Warping}: A disc that is not flat but does not show signs of breaking. Here the angular momentum profile varies as a function of radius, i.e. $\psi(R) \ne 0$, but there is no indication of minima in the surface density profile.
    \item \emph{Unsuccessful breaking}: A disc that shows the characteristics of breaking but does not actually separate into distinct smaller discs. The warp profile shows a local maximum with a corresponding local minimum in the surface density profile. However, as the disc evolves, $\psi_{\rm max}$ may increase but $\Sigma_{\rm min}$ does not continue to decrease.
    \item \emph{Successful breaking (single)}: A disc that shows the characteristics of breaking with an increasing $\psi_{\rm max}$ and decreasing $\Sigma_{\rm min}$, successfully separating into two smaller discs. In these simulations, $\Sigma_{\rm min}$ continues to decrease until $\Sigma_{\rm min}\approx 0$.
    \item \emph{Successful breaking (multiple)}: A disc that shows the characteristics of breaking at many radii simultaneously. The warp profile has local maxima and the surface density approaches zero at multiple locations, leading to several rings tearing off the disc.
\end{enumerate}

As shown in Fig.~\ref{fig:tearing_example}, the condition $\psi_{\rm max}$ and $\Sigma_{\rm min}$ are strongly correlated. In the case that the disc is stable, there is no or a weak $\psi_{\rm max}$ and no $\Sigma_{\rm min}$. As this disc evolves $\psi_{\rm max}$ decreases, making the disc more stable against breaking (see Sec.~\ref{section:longterm} for further discussion). In the case that the disc is unstable, the disk starts to break when both the surface density decreases and the warp profile increases at the same radial location. The disc then actually separates if this process continues until $\Sigma_{\rm min} \approx 0$. Previous 1D analyses  \citep{2015MNRAS.449.1251D,2018MNRAS.476.1519D,2020MNRAS.496.3060G} could only perform a coarser distinction between connected and disconnected discs. On the other hand, the 3D hydrodynamical simulations presented in this paper allow us to delineate between unsuccessful, single, and multiple breaking.

\subsection{Influence of the binary companion}

Let us consider a BH of mass $M$ and spin $J = GM^2 \chi/c$ (where $\chi\in[0,1]$ is the Kerr parameter) embedded in a disc with \cite{1973A&A....24..337S} viscosity $\alpha$. In the absence of a binary companion, the disc angular-momentum profile with radius is self-similar: BHs with different masses and spins will be surrounded by re-scaled versions of the same gaseous structure \citep{1996MNRAS.282..291S,2007MNRAS.381.1617M}.

If instead the BH is part of a binary system, the disc is subjected to both the Lense-Thirring torque at small radii as well as the tidal torque from the companion at large radii. In this case the dynamics depends on the interaction between the two external torques in the disc. Each torque has an associated radius where it creates the largest warp; at $R_{\rm LT}$ the Lense-Thirring torque most strongly affects the warp profile while at $R_{\rm tid}$ the tidal torque does the same. Following \citet{2020MNRAS.496.3060G}, we parameterise the interaction of the binary by introducing the non-dimensional `companion parameter' $\kappa$ with
\begin{align}
    \kappa = \left( \frac{R_{\rm tid}}{R_{\rm LT}} \right)^{-7/2}\,.
\end{align}
We set $R_{\rm LT}$ to be the reference radius such that $R_{\rm LT} = R_0$. The advantage of this parameterisation is that it encodes information on the companion's mass $M_\star$ and orbital separation $R_\star$. In the notation of \citet{2020MNRAS.496.3060G}, this `companion parameter' reads 
\begin{align}
\kappa
&\simeq 0.66
\left(\frac{M}{10^7 M_\odot} \right)^2
\left(\frac{\chi}{0.5} \right)^{2} 
\left(\frac{M_\star}{10^7 M_\odot} \right)
\left(\frac{R_\star}{0.1 {\rm pc}} \right)^{-3}
\notag \\&\;\;{\times
\left(\frac{H_0/R_0}{0.002} \right)^{-6}
\left(\frac{\alpha}{0.2} \right)^{-3}
\left[\frac{\zeta}{1/(2\!\times\! 0.2^2)} \right]^{-3}\,,}
\label{equation:kappa}
\end{align}
where
\begin{align}
    \zeta = \frac{2(1+7\alpha^2)}{\alpha^2(4 + \alpha^2)}\,.
\end{align}
where in the limit of $\alpha \rightarrow 0$, $\zeta \rightarrow 1/(2\alpha)$ and $\kappa \propto \alpha^3$. Here $H_0/R_0$ is the aspect ratio set at radius $R_0$ where the Lense-Thirring torque most strongly warps the disc. From  Eqs. (16) and (20) in  \citet{2020MNRAS.496.3060G}, this radius is equivalent to
\begin{align}
    R_0 = \frac{GM}{c^2} \left( \frac{H_0}{R_0} \right)^{-4/3} \left( \frac{4\chi}{\alpha \zeta}\right)^{2/3}\,.
    \label{equation:R_LT}
\end{align}

The case $\kappa = 0$ corresponds to the self-similar case of a single BH, while larger values of $\kappa$ correspond to configurations where the companion BH strongly perturbs the disc evolution. This could be due to the binary being massive (large $M,M_\star$), the BH being more rapidly rotating (large $\chi$),  the orbit being tight (small $R_\star$), the disc being thin (small $H_0/R_0$), a low viscosity (small $\alpha$),  or some combination of these. For binaries inspiralling under gas-assisted migration, the parameter $\kappa$ increases with time and may eventually lead the system to criticality.

\subsection{Break radius}
If the disc breaks, this is likely to happen at a radius $R_{\rm break}$ that maximizes the warp profile $\psi$.  \cite{2009MNRAS.400..383M} estimated this location by matching the external torques due to Lense-Thirring precession and the companion to find
\begin{align}
    R_{\rm break} = \left(\frac{8 G^{1/2} M^{1/2} J R_\star^3}{3 c^2 M_\star}\right)^{2/9}\,.
    \label{equation:tear_radius_companion}
\end{align}

In particular, Eq.~(\ref{equation:tear_radius_companion}) is independent of the disc viscosity $\alpha$. Therefore, we do not expect  significant differences in the location of the breaking radius if the disc is in the diffusive or wave-like regime. However, we note that this prediction relies solely on the balance between Lense-Thirring and companion tidal torques and neglects, by definition, the response of the disc to the warp propagation. Because of this, Eq.~(\ref{equation:tear_radius_companion}) does not (somewhat unphysically) depend on the relative inclination between the BH and the companion's orbit. As described below, the relative inclination does play a role in the breaking dynamics (see also  Fig.~\ref{fig:tearing_example}).

\subsection{Disc backreaction}
\label{backr}
As the BH warps the disc, the disc reacts by aligning the BH with its own angular momentum \citep{1976IAUS...73..225R}. The  evolution of the angle $\theta$ between the BH spin $\mathbfit{J}$ and the binary's orbital angular momentum $\mathbfit{L}_*$ is given by 
\begin{align}
    \frac{\rm d \cos \theta}{\rm d t} = 
    \frac{\rm d \hat{\mathbfit{J}}}{\rm d t} \cdot \hat{\mathbfit{L}}_*\,.
    \label{equation:dcthetadt}
\end{align}
The change of angular momentum of the BH $\rm d \mathbfit{J}/dt$ can be calculated from the integral of the torque exerted by the disc onto the BH \citep{2020MNRAS.496.3060G}
\begin{align}
    \frac{\rm d \mathbfit{J}}{dt} = -\int_{R_{\rm min}}^{R_{\rm max}}
        \frac{2G}{c^2} \frac{\mathbfit{J} \times \mathbfit{L}}{R^3}
        2\pi R \rm d R\,,
    \label{equation:dJdt}
\end{align}
where $\mathbfit{L}$ is the angular momentum of the disc, $\abs{\mathbfit{L}} = \Sigma \sqrt{GMR}$ and $R_{\rm min}$ and $R_{\rm max}$ are the inner and outer extent of the disc.

Ideally, one would like to fully take into account the back-reaction of the disc onto the BH  and evolve the system self-consistently. This was possible with the 1D scheme of \cite{2020MNRAS.496.3060G} but they had to rely on a quasi-adiabatic treatment and could only follow the evolution up to the critical obliquity. Although tracking the binary inspiral is prohibitive for our 3D simulations, one can still use the above description to investigate how the spin aligns on short timescales while the disc reaches its steady state. This is measured from our simulations by assuming azimuthal asymmetry to evaluate Eq.~(\ref{equation:dJdt}) and (\ref{equation:dcthetadt}) for each  simulation snapshot (see Appendix~\ref{section:calculating_dcdt} for a detailed explanation).

\section{Numerical simulations}
\label{secnumerics}

Our simulations are performed with the 3D SPH code \textsc{Phantom} \citep{2018PASA...35...31P}. This code has been used extensively to model inclined discs around BHs \citep{2013MNRAS.434.1946N,2015MNRAS.448.1526N}, to examine disc breaking \citep{2012ApJ...757L..24N,2015MNRAS.449.1251D}, and for comparison with analytical predictions of warped discs \citep{2010MNRAS.405.1212L}.

We conduct a total of 143 simulations considering multiple viscosities, binary separations, disc aspect ratios and inclinations. Here we detail the relevant aspects of the code for our application and the initial conditions used in all of our runs. Additionally, to the best of our knowledge a portion of our parameter suite constitutes the thinnest discs around BHs simulated to date with SPH (albeit marginally). Simulations are run until they show successful breaking or 15 binary orbits. Although this is a short time compared to the viscous time of the disc, the entirety of 143 simulations required $\sim 2.5$ million CPU hours (we discuss this limitation in Section~\ref{section:longterm}).

The vast majority of the discs in our parameter suite are comfortably in the `diffusive' regime, where $\alpha \gtrsim H/R$ and the warp propagates diffusively \citep{1983MNRAS.202.1181P}. For the discs that have $\alpha \sim H/R$, we may expect tilt oscillations to occur as seen by \citet{1997MNRAS.285..394I,2002MNRAS.337..706L} and \citet{2015MNRAS.448.1526N}. However, our discs are often not simulated until they reach a steady state and, as we will discuss next, we set a slightly larger outer boundary than was considered in these previous works. We thus do not expect (nor do we recover) evidence of tilt oscillations here.

\subsection{Black-hole modeling}
\label{section:modelling_BH}
In \textsc{Phantom} the rotating BH is modelled using a fixed potential. To do this we use a post-Newtonian approximation to model the potential of the rotating BH, achieved with a first order in $v/c$ correction in the momentum equation \citep{2000MNRAS.315..570N}
\begin{align}
    \frac{\text{d} \bm{v}}{\text{d} t} = -\frac{1}{\rho} \nabla P + \bm{v} \times \bm{h} - \nabla \Phi + S_{\rm visc}\,,
    \label{equation:momentum}
\end{align}
where $\bm{v}$, $\rho$ and $P$ are the gas velocity, density and pressure respectively and $S_{\rm visc}$ is the viscous force per unit mass. The term $\bm{v} \times \bm{h}$ represents the gravitomagnetic force per unit mass, where
\begin{align}
        \bm{h} = \frac{2\chi G^2M^2}{R^3c^3} \left(\bm{\hat{J}} - 3 \frac{(\bm{\hat{J}}\cdot\bm{r})\bm{r}}{R^2} \right)\,,
        \label{equation:h}
\end{align}
$\mathbfit{r}$ is the spherical coordinate vector, and $R$ is the distance to the primary BH. We use the modified potential  of \cite{2000MNRAS.315..570N}
\begin{align}
    \Phi(R) = -\frac{GM}{R} \left(1 + \frac{3R_{\rm g}}{R} \right)\,,
\end{align}
where $R_{\rm g}=GM/c^2$ is the gravitational radius of the BH. While preventing the gravitational force from tending to infinity close to the BH, this expression  also accurately recovers both apsidal precession frequency at large radii and the sign of the nodal precession frequency. We note that the modified potential looses accuracy for $R\lesssim 10R_{\rm g}$ but this is within our numerical accretion radius.

We note that using a fixed potential limits our simulations, in that physically it is equivalent to assuming that the primary BH is located at the centre of mass of the binary system. To respect this assumption, we only consider systems with a mass ratio $M_\star / M \approx 0.01$ and a low disc mass such that the centre of mass of the system is within the accretion radius of the BH (see Sec.~\ref{section:ics}). We vary the other parameters entering Eq.~(\ref{equation:kappa}) to span a wide range of $\kappa$ values, above and below criticality.

\subsection{Initial conditions}
\label{section:ics}
Each of our simulations is initialised with a disc and binary companion in orbit around the primary BH.  Here we detail the properties of the initial conditions, noting that simulations are non-dimensionalised by using $\kappa$ from Eq.~(\ref{equation:kappa}). 
Our parameters and the corresponding values of $\kappa$ are summarised in Table~\ref{tab:initial_conditions}.

\begin{table*}
\renewcommand{\arraystretch}{1.3}
\setlength{\tabcolsep}{6pt}

 \begin{tabular*}{0.7\textwidth}{ccccccccc} 
  \hline
  $\kappa$ & $R_{\star}/R_{\rm g}$ & $R_{\rm out}/R_{\rm g}$ & $\alpha$ & $H_{\rm ref}/R_{\rm ref}$ & $R_0/R_{\rm g}$ & $H_0/R_0$ & $\theta$ & $N$\\
  \hline
  \hline
  $3.9 \!\times\! 10^{-5}$ & 398 & 250 & 0.10 & 0.08 & 19.2 & 0.089 & 20$\degr$--160$\degr$ & 14\\
  $1.4 \!\times\! 10^{-4}$ & 398 & 250 & 0.05 & 0.05 & 25.9 & 0.052 & 20$\degr$--160$\degr$ & 12\\
  $1.7 \!\times\! 10^{-4}$ & 398 & 250 & 0.15 & 0.08 & 26.8 & 0.082 & 20$\degr$--160$\degr$ & 15\\
  $5.9 \!\times\! 10^{-4}$ & 250 & 150 & 0.05 & 0.05 & 25.9 & 0.052 & 10$\degr$--160$\degr$ & 10\\
  $2.7 \!\times\! 10^{-3}$ & 398 & 150 & 0.10 & 0.05 & 49.3 & 0.044 & 20$\degr$--160$\degr$ & 13\\
  $1.2 \!\times\! 10^{-2}$ & 398 & 250 & 0.15 & 0.05 & 68.5 & 0.041 & 20$\degr$--160$\degr$ & 11\\
  $2.8 \!\times\! 10^{-2}$ & 398 & 250 & 0.20 & 0.05 & 83.0 & 0.039 & 20$\degr$--160$\degr$ & 10\\
  $2.6 \!\times\! 10^{-1}$ & 398 & 250 & 0.10 & 0.03 & 136.8 & 0.021 & 20$\degr$--160$\degr$ & 14\\
  $1.2$ & 398 & 250 & 0.15 & 0.03 & 190.3 & 0.019 & 20$\degr$--160$\degr$ & 14\\
  \hline
  $2.1 \!\times\! 10^4$ & 250 & 150 & 0.10 & 0.01 & $1.2 \times 10^3$ & 0.0040 & 20$\degr$--160$\degr$ & 11\\
  $9.2 \!\times\! 10^4$ & 250 & 150 & 0.15 & 0.01 & $1.7 \times 10^3$ & 0.0036 & 20$\degr$--160$\degr$ & 10\\
  $2.2 \!\times\! 10^5$ & 250 & 150 & 0.20 & 0.01 & $2.1 \times 10^3$ & 0.0034 & 20$\degr$--160$\degr$ & 9\\ 
  \hline
 \end{tabular*}
  \caption{Summary of the initial conditions of our simulations, ordered by increasing $\kappa$. Here $\kappa$ is a measure of the relative influence of the secondary [cf. Eq.~(\ref{equation:kappa})], $R_\star$ is the binary separation, $R_{\rm out}$ is the outer radius of the disc, $\alpha$ is the disc viscosity, $H_{\rm ref}/R_{\rm ref}$ the disc aspect ratio at $R_{\rm ref}$, $R_0$ is the Lense-Thirring radius [cf. Eq.~(\ref{equation:R_LT})], $H_0/R_0$ is the disc aspect ratio at $R_0$, $\theta$ is the initial inclination between the disc and BH, and $N$ is the number of simulations we have completed in the listed $\theta$ range. The simulations in the upper (lower) part of the table are presented in Fig.~\ref{fig:break_notbreak} (Fig.~\ref{fig:high_kappa}).} \label{tab:initial_conditions}
\end{table*}

The disc is initialised as a flat disc in the $x$-$y$ plane (i.e., in the plane of the binary orbit) with a mass of $10^{-6} M$. This disc mass is deliberately low to respect our assumption of using a fixed potential and to avoid any back-reaction effect on the properties of the binary companion. The surface density profile is given by
\begin{align}
    \Sigma(R) = \Sigma_{\rm ref} \left(\frac{R}{R_{\rm ref}} \right)^{-1} \left(1 - \sqrt{\frac{R_{\rm in}}{R}} \right)\,,
\end{align}
where the normalisation $\Sigma_{\rm ref}$ is determined from the disc mass, $R_{\rm ref}=30R_{\rm g}$ is the reference radius, and $R_{\rm in}=15R_{\rm g}$ is the inner edge of the disc. The outer radius of the disc $R_{\rm out}$ is set to either $150R_{\rm g}$ or $250R_{\rm g}$ depending on the orbit of the  companion (cf. Table~\ref{tab:initial_conditions}). Our results are reported in orbits of the binary companion, with all simulations running for a minimum of 3 orbits of the binary or $\sim$150 orbits at $R_{\rm ref}$.

We assume that the disc is vertically isothermal, such that the sound speed in the disc can be described by $c_{\rm s}(R) = c_{\rm s,ref} (R/R_{\rm ref})^{-q}$ with $q=3/4$. Here $c_{\rm s,ref}$ is determined by the disc thickness (aspect ratio), with $(H_{\rm ref}/R_{\rm ref}) = 0.01, 0.03, 0.05$ and $0.08$ set at the reference radius $R_{\rm ref}$ (cf. Table~\ref{tab:initial_conditions}). While these values are relatively large for AGN discs \citep{2009ApJ...700.1952H}, we are limited by our numerical resolution to $H/R\gtrsim 0.01$. We stress that the dynamics only depend on the companion parameter $\kappa$: results obtained with larger aspect ratio will still be robust for discs with a lower $H/R$ but the same value of  $\kappa$.

We model the viscosity in the disc using the \citet{1973A&A....24..337S} prescription, with $\alpha = 0.05, 0.10, 0.15$ and $0.20$.  This is implemented in \textsc{Phantom} using the shock viscosity term described by \citet{2018PASA...35...31P} \citep[but see also][]{1994ApJ...421..651A,1996MNRAS.279..402M,2010MNRAS.405.1212L}. The artificial viscosity coefficient $\alpha_{\rm AV}$ is related to the physical viscosity by
\begin{align}
    \alpha \approx \frac{1}{10} \alpha_{\rm AV} \frac{\langle h \rangle}{H}\,,
\end{align}
where $\langle h \rangle$ is the shell-averaged smoothing length. For a given resolution determined by $\langle h \rangle/H$, we set $\alpha_{\rm AV}$ to give the targeted $\alpha$. \citet{2018PASA...35...31P} suggests that $\alpha_{\rm AV} \gtrsim 0.1$ is necessary to resolve the physical viscosity and for all of our simulations we have $\alpha_{\rm AV} \gtrsim 0.85$. Our choice of surface density and sound speed power-law profiles implies that $\alpha$ varies with radius throughout the disc following a power-law with index  $-1/6$ \citep{2010MNRAS.405.1212L}. Once  simulations begin evolving, we find that $\langle h \rangle/H$ becomes roughly constant across the vast majority of the disc (except where breaking occurs), essentially removing any radial variation in $\alpha$.

In the simulations the BH has a spin of $\chi=0.9$ and the mass is set to $M=1$ (alongside $G=c=1$), noting that for our figures we rescale it to $M=10^7 M_\odot$. To accommodate a relative inclination between the disc and BH, the spin angular momentum vector of the BH is set to $\bm{\hat{J}} = (\sin \theta, 0, \cos \theta)$ where $\theta$ is the relative inclination. Thus a BH with $\theta=0$ would have spin along the $z$ axis relative to a disc that is initialised in the $x$-$y$ plane. Material that falls inside $R_{\rm in}$ is accreted without further checks.

We restrict the mass of the binary companion to be relatively low to accommodate our use of a fixed BH potential (see Sec.~\ref{section:modelling_BH}). For all of our simulations, the mass of the binary companion is set to $M_\star = 10^{-2} M$ and the semi-major axis $R_\star = 250, 398R_{\rm g}$ as in Table~\ref{tab:initial_conditions}. The binary companion is initially set on a circular orbit neglecting the disc mass in the $x$-$y$ plane. The accretion radius of the companion is set to $0.25$ of the Hill radius, equivalent to $14.6R_{\rm g}$ for $R_\star=398R_{\rm g}$ and $9.2R_{\rm g}$ for $R_\star=250R_{\rm g}$, in line with previous guidance by \cite{2018MNRAS.481...20N}. Although the orbit of the companion is free to evolve and feels the back-reaction from the disc, this effect is negligible due to the low mass ratio ($M_{\rm disc}/M_\star=10^{-4}$).

Accurately resolving the disc is crucial to recovering disc breaking \citep{2015MNRAS.448.1526N}. We use $N=5 \times 10^6$ particles for each simulation, which ensures that the discs are well resolved. To check this we measure the average smoothing length to disc scale height ratio $\langle h \rangle/H$ as in \citet{2010MNRAS.405.1212L}. The disc is discretised into 300 radial annuli and the particle properties in each annuli averaged to produce radial profiles. To take into account the warping of the disc, we additionally measure the disc scale height from the instantaneous warped mid-plane. We meet the resolution criteria $\langle h \rangle/H < 0.5$ in all but the innermost region of the disc across all of our simulations.

\subsection{Evaluating the companion parameter $\kappa$}
In order to scale our simulations consistently with the description of \citet{2020MNRAS.496.3060G}, we need to evaluate $\kappa$ which in turn depends on both $R_0$ and $H_0/R_0$. To connect these parameters, we use the analytic expression of the aspect ratio
\begin{align}
\label{HRpowlaw}
    \frac{H}{R}=  \frac{H_{\rm ref} }{R_{\rm ref} } \left( \frac{R}{R_{\rm ref}} \right)^{1/2 - q}\,,
\end{align}
and Eq.~(\ref{equation:R_LT}), solving for $R_0$ in terms of $R_{\rm g}$. This yields
\begin{align}
    R_0 = \left( \frac{4 \chi}{\alpha \zeta} \right) \left(\frac{H_{\rm ref} }{R_{\rm ref} }\right)^{-2} \left(\frac{R_{\rm ref}}{R_g}\right)^{-1/2} R_{\rm g}\,,
\end{align}
where $(H_{\rm ref}/R_{\rm ref})$ is the aspect ratio at the reference radius in our simulation initial conditions. From $R_0$, one can then evaluate $H_0/R_0$ from  Eq.~(\ref{HRpowlaw}) and thus $\kappa$ from Eq.~(\ref{equation:kappa}). The resulting values are reported in Table~\ref{tab:initial_conditions}.

\subsection{Quantifying a breaking disc}
\label{section:diditbreak}
Given the size of our parameter suite, we desire an automated process to identify if and where any given simulation exhibits tearing. Based on the criteria highlighted  in Sec.~\ref{section:defining_a_break}, we consider either unsuccessful or successful tearing to occur when we find a radially correlated local minimum in $\Sigma$ and local maximum in $\psi$ ---that is, where a given $\Sigma_{\rm min}$ and $\psi_{\rm max}$ occur at the same $R$. In particular, we consider these to be at the same radii when they are within three radial bins of each other, corresponding to 1\% of the full radial domain. Successful tearing occurs when additionally $\Sigma_{\rm min} \approx 0$, which we accept when $\Sigma$ drops below 10\% of the maximum of the initial surface density profile.

Our discs feature spiral arms at large radii due to the tidal interaction with the binary. Crucially, these can be picked up as false positives by the above criteria. We thus also require that $\psi_{\rm max} > 1.0$ to confirm tearing, as most the spirals tend to be associated with values of $\psi_{\rm max}$ that are considerably lower. All borderline cases were also visually inspected. We use this procedure across our simulations for all time-steps to identify and locate the breaking radius.

\begin{figure*}
    \centering
    \includegraphics[width=0.83\textwidth]{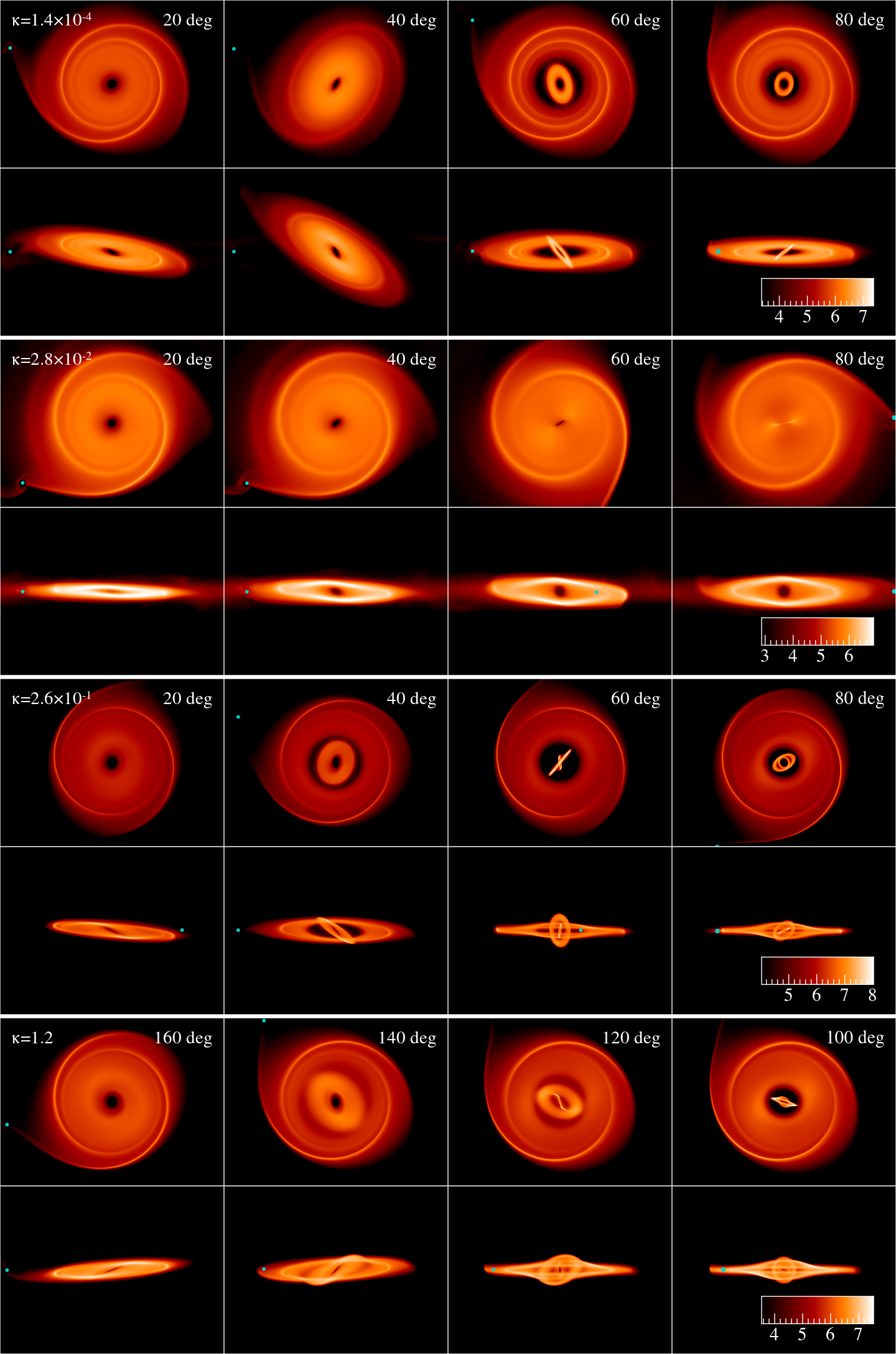}
    \caption{Density renderings for a representative set of our suite of 143 simulations showing examples of warping, unsuccessful breaking and successful breaking. Common to all simulations are the spiral arms at the outer disc edge driven by the tidal interaction with the BH companion (shown in green). Top to bottom the view alternates between the $x$-$y$ and the $x$-$z$ plane. The colour scale shows column density in $\rm kg / \rm m^2$ and all panels are shown on the same spatial scale. Simulations are shown at their end time (different for each). A summary of the full suite is shown in Fig.~\ref{fig:break_notbreak}.}
    \label{fig:composite}
\end{figure*}

\section{Results}
\label{secresults}

\subsection{Qualitative behaviour}

Figure~\ref{fig:composite} shows a demonstrative selection of the discs from our simulated suite. Broadly speaking, all of our simulations evolve similarly, with a warp developing in the inner regions while the outer disc is shaped by the tidal interaction with the binary companion.

Within the first couple of orbits, the binary companion induces two spiral arms which tends to be more pronounced in the simulations with the smaller aspect ratio.
As expected, the structure of the spiral arms is not affected by the relative inclination of the primary BH and are sustained throughout the duration of the simulations (irrespective of whether the disc tears or not). Additionally, the orbit of the binary does become inclined slightly, but this is a small effect ($\lesssim 1 \degr$).

The inner disc evolution depends on whether the disc warps, breaks into two sections, or breaks into multiple rings (cf. Fig.~\ref{fig:tearing_example}). In all cases, the inner region ($R\lesssim 50R_{\rm g}$) shows evidence of a warp within the first couple of orbits of the binary companion. As the disc continues to evolve, numerous discs in our suite show visual evidence of break. For those that show multiple rings, as in previous work we find that the rings tear off successively from the inner region outwards \citep[e.g.][]{2013MNRAS.434.1946N}. The thickness of the sections that tear off correspond to the disc thickness, with thinner rings forming for those discs with the smaller aspect ratio. The broken components precess differentially resulting in a range of misalignments relative to the outer disc. In a handful of simulations, the broken ring or disc also exhibits some local asymmetric perturbations but these are short lived.

\begin{figure*}
    \centering
    \includegraphics[width=2\columnwidth]{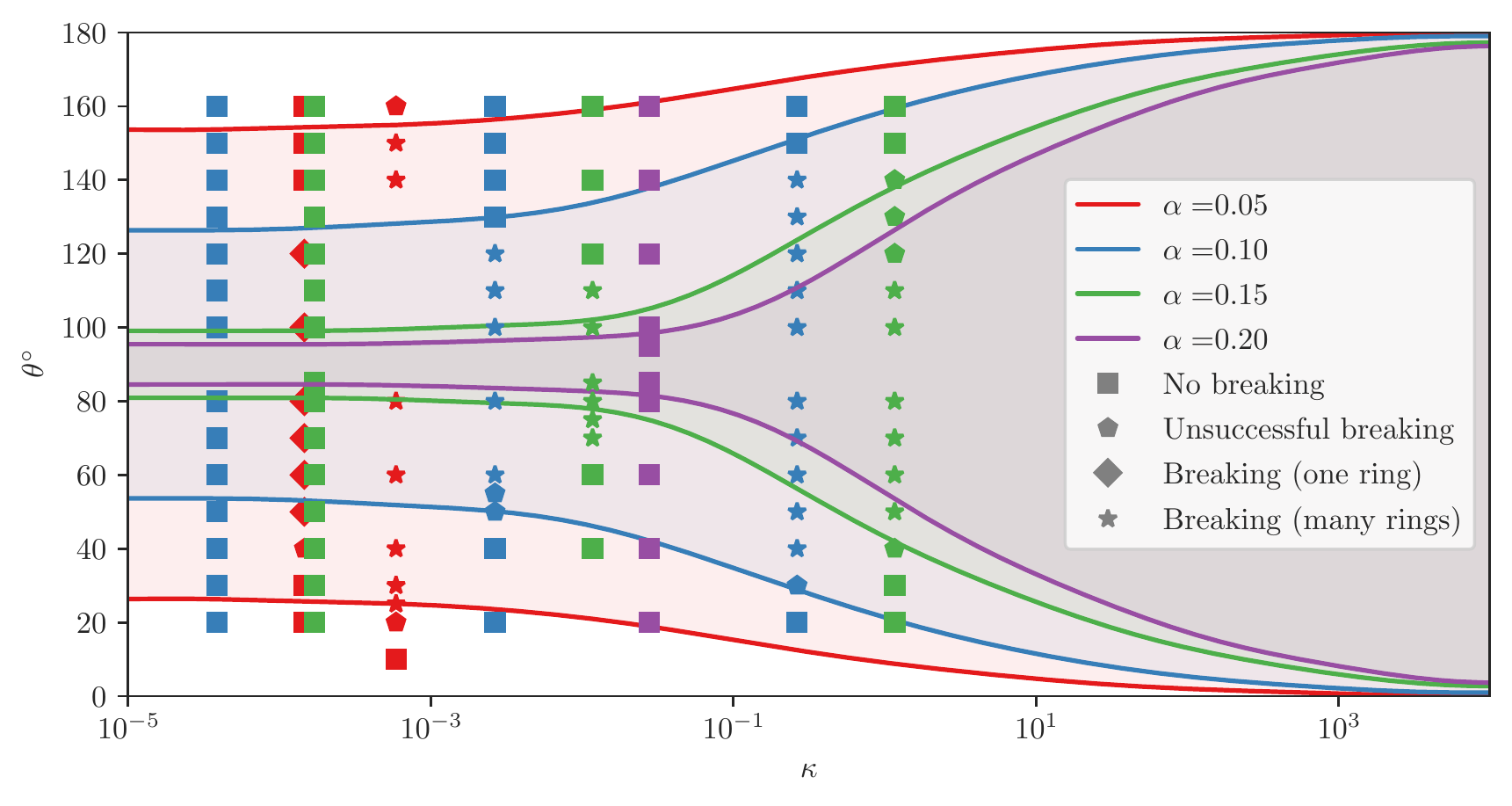}
    \caption{Comparison of our 3D hydrodynamical simulations to the 1D semi-analytic prediction by \citet{2020MNRAS.496.3060G} as a function of $\kappa$ (Eq.~\ref{equation:kappa}) and the initial BH misalignment $\theta$. Here the behaviour of the disc in our simulations is coded with warped discs as squares, unsuccessful breaking as pentagons, successful breaking into two discs with diamonds, and tearing into many rings as stars. The shaded regions indicate the critical region identified by \citet{2020MNRAS.496.3060G} where 1D disc solutions cannot be found. Colours represent the viscosity of the discs.}
    \label{fig:break_notbreak}
\end{figure*}

By the end of our simulations the entire disc has tilted away from its initial orientation in the $x$-$y$ plane, but the final inclination depends on whether the disc has broken or not. In the cases where the disc has not broken, the outer edge of the disc is almost co-planar with the orbit of the binary and the inner region is warped. Alternatively, when the disc tears, the outer disc has an inclination that lies between that of the binary companion and the equatorial plane of the BH, with the inner broken regions more strongly misaligned (for example, see the lower row of Fig.~\ref{fig:composite}).

\subsection{Testing the semi-analytic predictions}
Figure~\ref{fig:break_notbreak} summarises our results for the different $\kappa$ values and relative inclinations $\theta$. Overplotted on Fig.~\ref{fig:break_notbreak} are the semi-analytic predictions by \citet{2020MNRAS.496.3060G} for where disc breaking should occur for each of the $\alpha$ values considered. Here we have used their $\beta=3/2$ solutions (in their notation, $\beta$ is the power-law index of the viscosity profile) but note that this choice does not alter our conclusions. The solid lines indicate the locations of the critical obliquity: for each value of $\alpha$, 1D solutions cannot be found in the central shaded region of the plot bounded by the two solid lines.

Overall, we find very good agreement between the 1D analytic prediction from \citet{2020MNRAS.496.3060G} and the results of our 3D hydrodynamic simulations. In the region where \citet{2020MNRAS.496.3060G} predicts breaking, we additionally distinguish cases of unsuccessful tearing, successful breaking and breaking with single vs. multiple rings. Far from criticality (i.e. above and below the solid lines), simulated discs tend to warp without  breaking (squares in  Fig.~\ref{fig:break_notbreak}). The semi-analytic  approach correctly describes the transition to a different regime characterised by either unsuccessful (pentagons) or breaking (diamonds or stars). Once the inclination is greater than criticality, we find that whether discs break into two or many discs depends on their disc properties and not their relative inclination.

Overall, we thus confirm that the critical obliquity does indeed correspond to disc breaking. 

\subsubsection{Large aspect ratios, $H/R$}
We do identify some cases where the 3D and 1D results differ substantially. Across our full parameter suite we do not observe breaking in any of the discs with $H/R=0.08$. This  suggests disagreement with the semi-analytic model, in particular for the discs with $\kappa=3.9\times10^{-5}$ and $1.7\times10^{-4}$ and lower inclinations. This discrepancy is unlikely to be due to the $\kappa$ and $\alpha$ values used, as the $\kappa=1.4\times10^{-4}$ series has a similar $\kappa$ and a lower $\alpha$ but good agreement. We also dismiss any resolution effects because these discs are thicker than the rest of our suite and thus present slightly better numerical resolution than the thinner discs. For these disc parameters, the simulations are also comfortably in the diffusive regime so we can safely disregard any potential issues due to wave-like behaviour. We thus conclude that the semi-analytic calculation, which is inherently designed to a model a disc-like structure, has an additional limitation in assuming that the disc is sufficiently thin. For the other parameters that we have held constant (i.e. $R_*$), this limitation corresponds to roughly $H/R \sim 0.08$. In support of this we find good agreement with the semi-analytic model for our $H/R=0.05$ discs and excellent agreement when $H/R=0.03$.

\subsubsection{Large companion parameters, $\kappa$}
\begin{figure}
    \centering
    \includegraphics[width=\columnwidth]{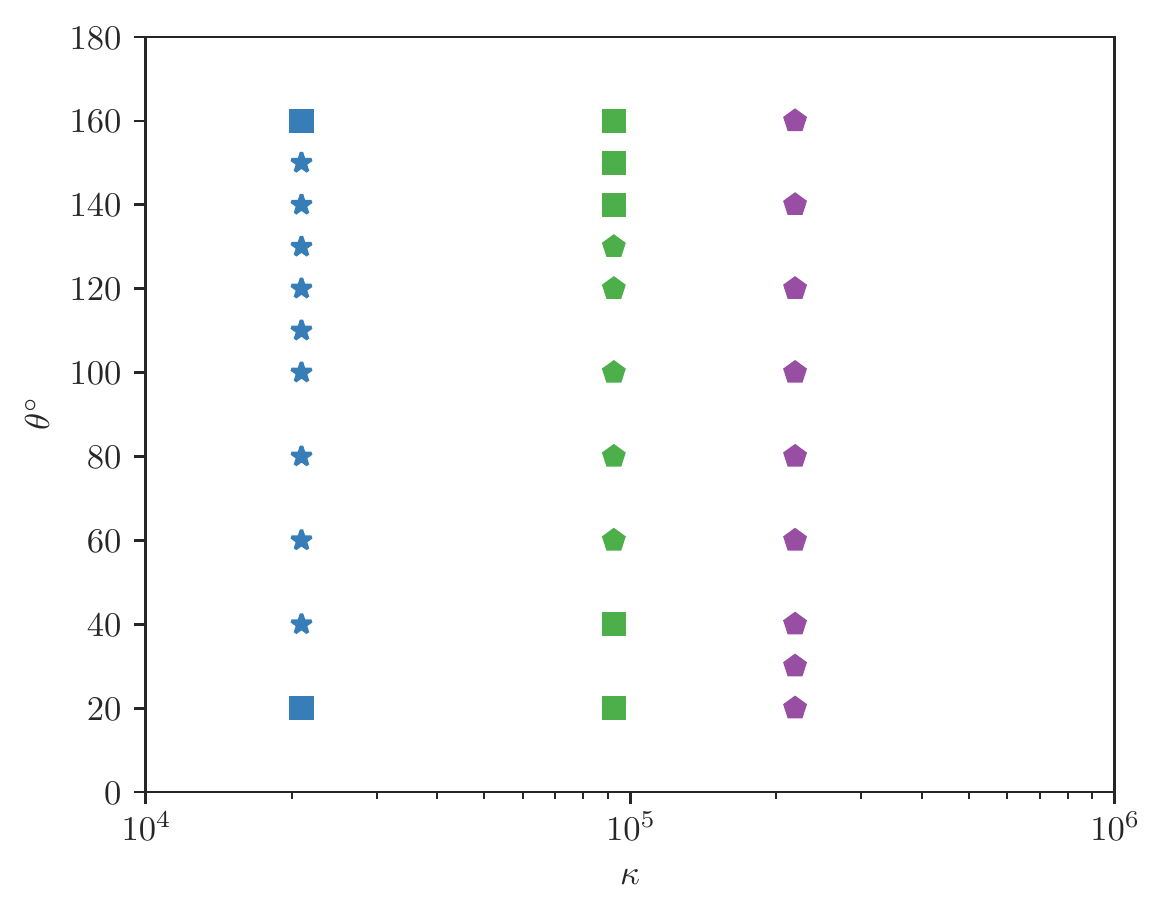}
    \caption{Same as Fig.~\ref{fig:break_notbreak} but for  additional simulation sets with higher $\kappa$ which violates the assumption of the semi-analytic 1D model. Even at these large $\kappa$ values We still find evidence of disc breaking in our 3D simulations.}
    \label{fig:high_kappa}
\end{figure}

We additionally run three sets of simulations at very large values of $\kappa > 10^4$, summarised in Fig.~\ref{fig:high_kappa}. Here the influence of the binary companion is much larger and our corresponding $R_0$ values are well outside the outer radius of the disc (c.f. Table~\ref{tab:initial_conditions}). Such large $R_0$ values are problematic when comparing to the semi-analytic model, as the current implementation of \citet{2020MNRAS.496.3060G} assumes that $R_0$ is located within the disc. For the runs presented earlier ($\kappa \lesssim 1.2$), the combination of the dimensionless scaling and radial range resulted in an inner boundary that was consistent with the inner boundary in our 3D simulations. If $\kappa$ is orders of magnitudes larger, however, their inner boundary falls well outside the inner boundary we have adopted. For example, with $\kappa=2.1\times10^4$ the inner boundary assumed by \citet{2020MNRAS.496.3060G} sits at $123R_{\rm g}$ which is almost at our simulated outer disc edge of $150R_{\rm g}$. A direct comparison between the two approaches is thus not possible for these large-$\kappa$, large-$R_0$ sets of simulations.

Despite this, at high $\kappa$ we find that Fig.~\ref{fig:high_kappa} still shows that for moderate to large inclinations we can expect breaking and for small inclinations, the disc to remain stable. We thus expect our predictions in relation to the BH disc alignment to be indicative even at higher $\kappa$ (see Section~\ref{section:disc_BH_alignment}).

\subsubsection{High viscosity, $\alpha$}
\begin{figure}
    \centering
    \includegraphics[width=\columnwidth]{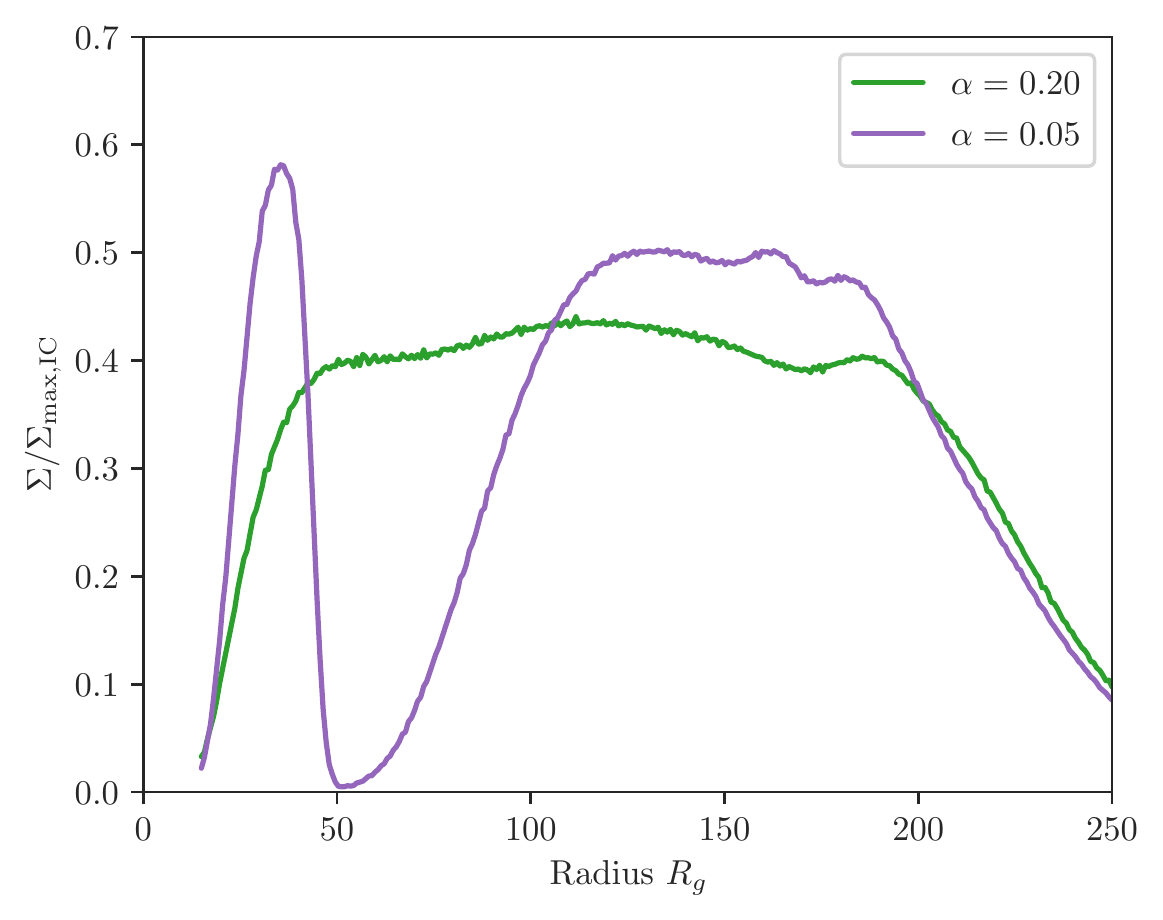}
    \caption{Surface density profiles showing that at high $\alpha$ the material where we might expect to see disc breaking is rapidly accreted. Here the simulations both have $\theta=80 \degr$ with $\alpha=0.05$, $\kappa=1.4 \times 10^{-4}$ and $\alpha=0.2$, $\kappa=2.8 \times 10^{-2}$. Both simulations are shown at 5.2 orbits.} 
    \label{fig:sigma_viscosity}
\end{figure}

We additionally have two sets of simulations that have $\alpha=0.2$, which at low $\kappa$ do not necessarily show agreement with the semi-analytic model. For the case with $\kappa=2.8 \times 10^{-2}$ and $\alpha=0.2$ this is particularly surprising because the semi-analytic model should work quite well at low $\kappa$ and high $\alpha$. As before, for these discs we can rule out any differences due to our numerical implementation as the cause of this discrepancy. However, consideration of the surface density profiles suggests that the inner regions --- where we expect breaking to occur --- is rapidly accreted. Figure~\ref{fig:sigma_viscosity} compares two $\Sigma$ profiles for simulations with $\alpha=0.05$ and $\alpha=0.2$. In the low $\alpha$ case the disc shows clear breaking at $\approx 50 R_{\rm g}$ and there is still a significant amount of material inside this radius making up the inner ring. The high $\alpha$ case has accreted much more material from the inner edge, potentially prohibiting the ring from breaking off.

\subsection{Location of the break radius}
\begin{figure}
    \centering
    \includegraphics[width=\columnwidth]{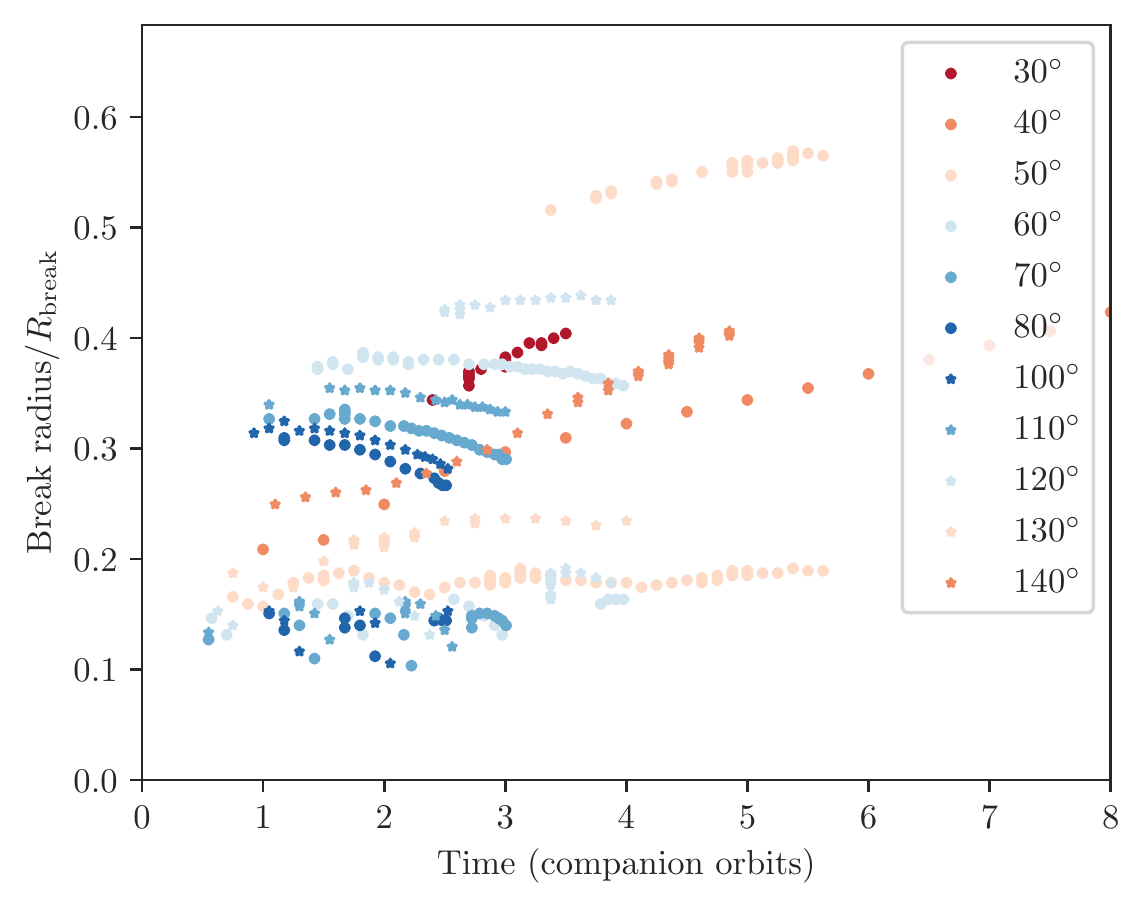}
    \caption{Location of the breaking radius measured in the simulations scaled by its predicted location, Eq.~(\ref{equation:tear_radius_companion}). We show our simulations with unsuccessful and successfully broken discs with $\kappa = 2.6 \times 10^{-1}$. In the early stages of these simulations there appears to be a dependence on the inclination. Retrograde discs (stars) consistently break at larger radii than their prograde counterparts (circles).}
    \label{fig:break_radius_location}
\end{figure}

Figure~\ref{fig:break_radius_location} shows a representative sample of the location of the breaking radius for our simulations. We find that, when breaking starts, the radius at which it occurs depends on the inclination but the relationship is unclear. The measured break radius is also systematically smaller than the prediction of Eq.~(\ref{equation:tear_radius_companion}). In most cases, the breaking radius decreases with time, although the opposite happens for a few runs (e.g. $\theta=40 \degr$, $140 \degr$ in Fig.~\ref{fig:break_radius_location}). Both of these observations are likely a consequence of the short duration of our simulations, as we only aim to demonstrate stability against breaking. Overall, we find that the disc breaks at a location that is up to $\sim 5$ times smaller than the prediction by \cite{2009MNRAS.400..383M}.

\begin{figure}
    \centering
    \includegraphics[width=\columnwidth]{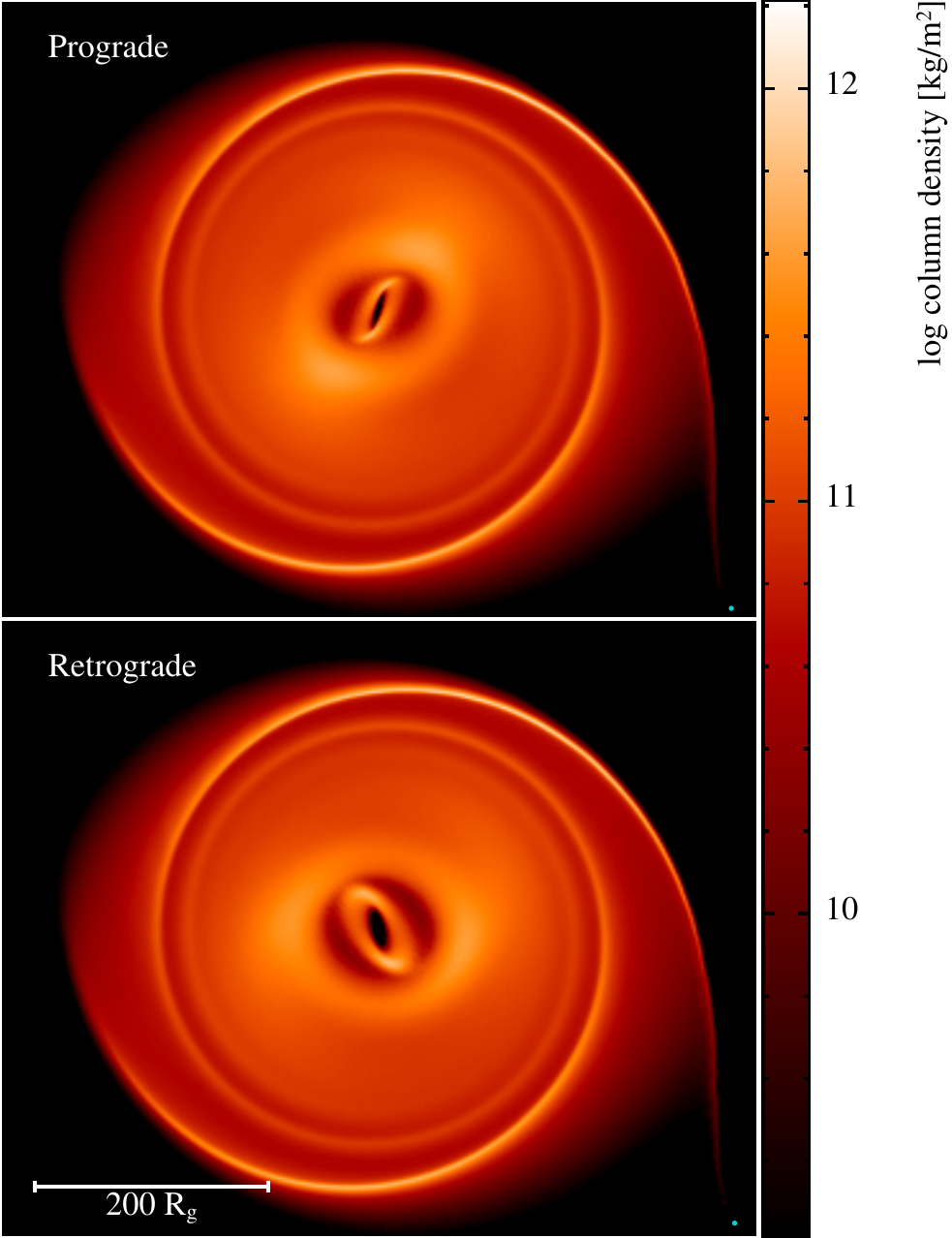}
    \caption{Comparing the density evolution of a prograde/retrograde pair of simulations after three companion orbits. While the 1D treatment implies perfect symmetry, in 3D simulations this is broken by the spirals induced by the binary companion. The prediction of whether the disc will break or not remains robust, see Fig.~\ref{fig:break_notbreak}, although these discs break at slightly different locations. The simulations shown here have $\kappa=1.45 \times 10^{-2}$ and $\theta=90\degr \pm 40\degr$.}
    \label{fig:pro_vs_ret_rendering}
\end{figure}

Across our parameter suite we consistently find that the break radius for retrograde discs is larger than for their prograde counterparts. Figure~\ref{fig:pro_vs_ret_rendering} shows an example prograde/retrograde pairing (with $\kappa=1.45 \times 10^{-2}$ and $\theta=90\degr \pm 40\degr$). The disc structure only differs in the inner regions with the retrograde case breaking at a slightly larger radius. \citet{2020MNRAS.496.3060G} finds that these cases are perfectly symmetric, with an identical dynamics. However, the direction of the spirals relative to the BH spin does depend on whether the BH is prograde or retrograde and thus breaks this assumption of symmetry. This an exclusively 3D effect that cannot be captured with simpler 1D models.  We confirm this behaviour for all of our prograde/retrograde pairings and note that, even though the details of the breaking are slightly different, the prediction of whether the disc will break or not is robust (Fig.~\ref{fig:break_notbreak}).

\begin{figure}
    \centering
    \includegraphics[width=\columnwidth]{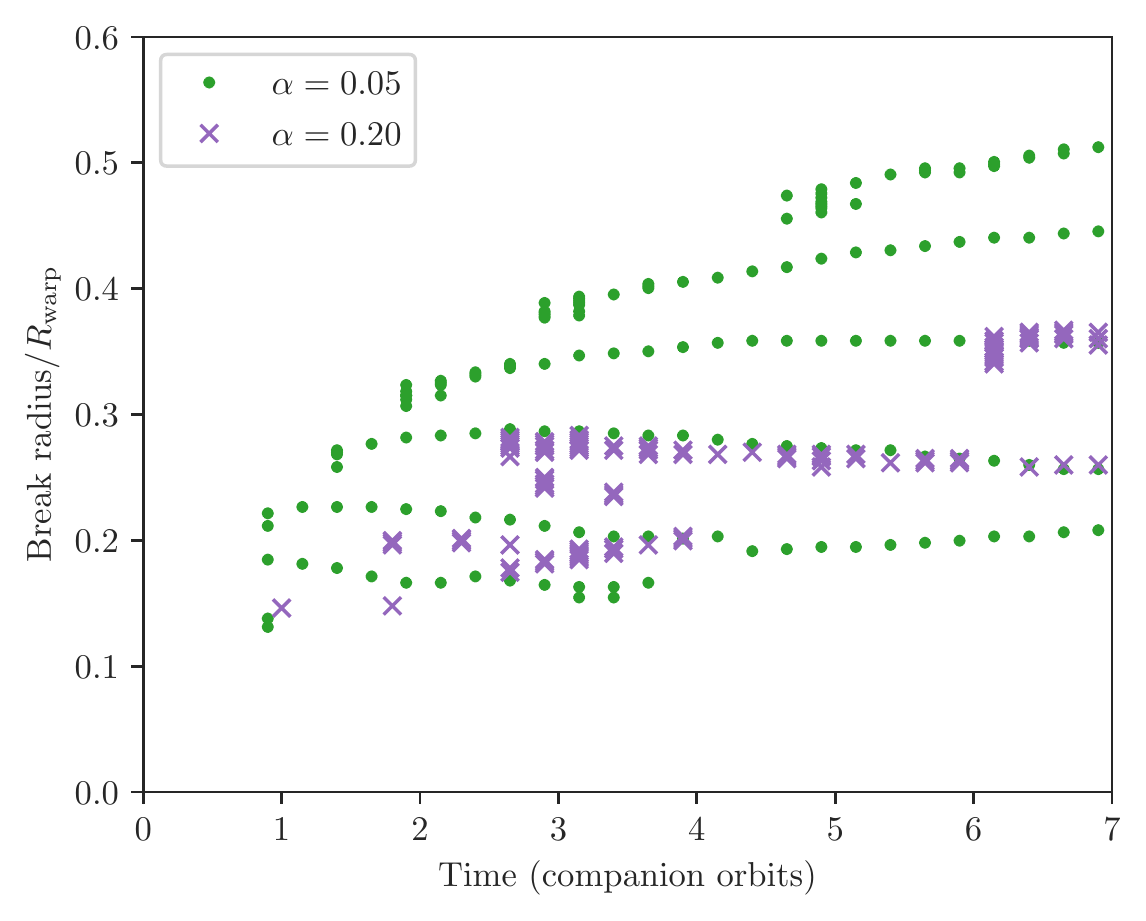}
    \caption{Location of signatures of tearing for simulations with the same BH and binary torque but different viscosities. Here the simulation with $\alpha=0.05$ has $\kappa=1.4\times10^{-4}$ and $\alpha=0.2$ has $\kappa=2.2\times10^5$. Although only the lower $\alpha$ disc successfully tears,  both show indications of tearing at the same radius, simultaneously confirming the predictions by \citet{2009MNRAS.400..383M} and \citet{2020MNRAS.496.3060G}.}
    \label{fig:dependence_on_viscosity}
\end{figure}

The predicted value of $R_{\rm break}$ in Eq.~(\ref{equation:tear_radius_companion}) is  independent of viscosity. In Fig.~\ref{fig:dependence_on_viscosity}, we compare two of our simulations with $\alpha=0.05$, $H/R=0.05$ and $\kappa=1.4\times10^{-4}$ vs $\alpha=0.2$, $H/R=0.01$ and $\kappa=2.2\times 10^5$. Here we hold $\theta=40 \degr$ and $R_*$ (and thus the external torques) as constant. The lower $\alpha$ case demonstrates tearing while the higher $\alpha$ case shows evidence of trying to tear but ultimately is unsuccessful, cf. Fig.~\ref{fig:break_notbreak}. This demonstrates that our simulations are consistent with predictions from both \citet{2009MNRAS.400..383M} and \citet{2020MNRAS.496.3060G}: while successful disc tearing depends strongly on the disc viscosity, the radius where it tears does not.

\subsection{Spiral arms can prevent disc tearing}
\label{section:spirals_stop_breaking}
\begin{figure}
    \centering
    \includegraphics[width=\columnwidth]{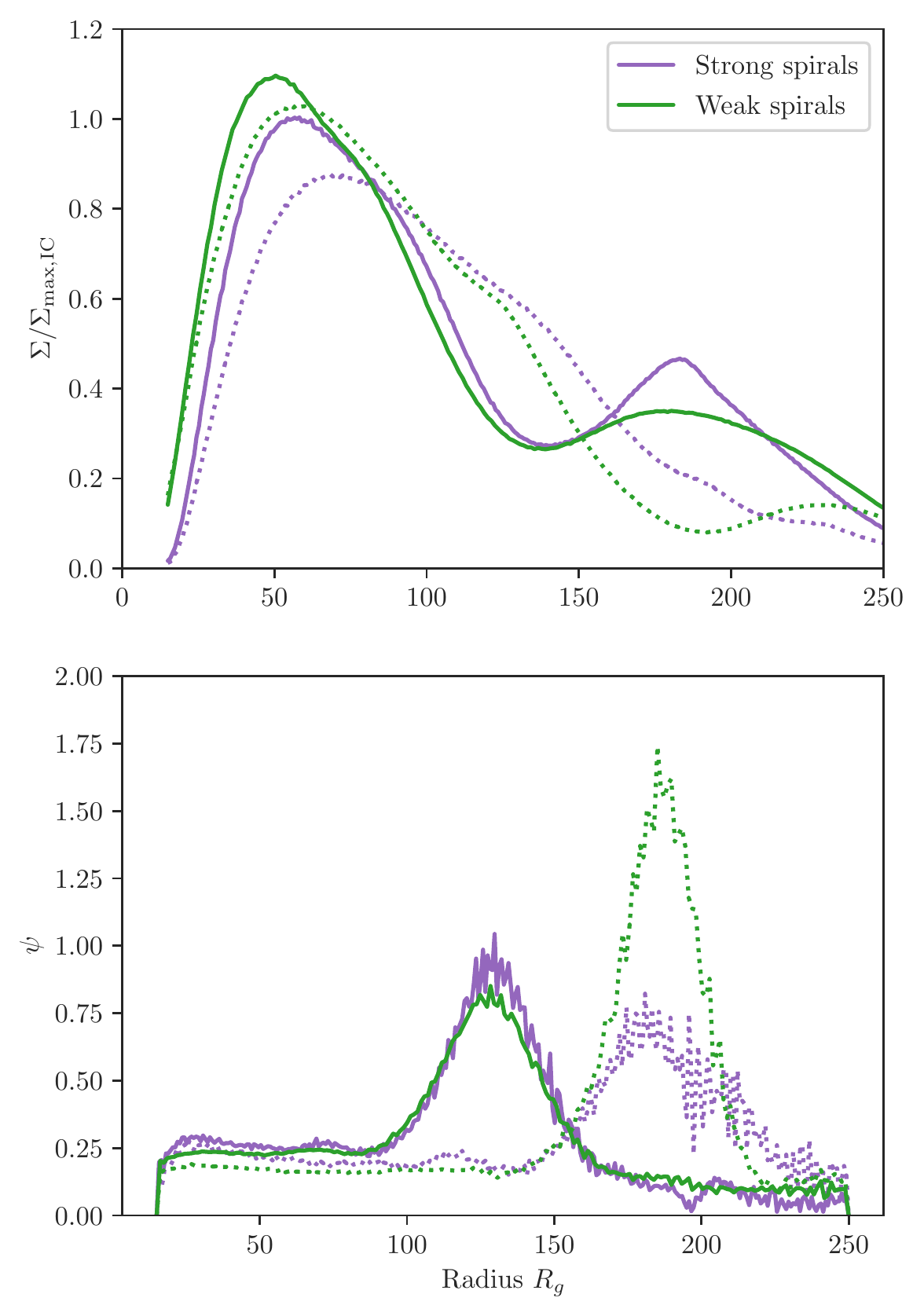}
    \caption{Surface density ($\Sigma$, top panel) and warp profiles ($\psi$, bottom panel) for discs with $\theta=40 \degr$, $\kappa = 1.4 \times 10^{-4}$ (`Strong spirals', purple) and $\theta=40 \degr$, $4.3\times 10^{-5}$ (`Weak spirals', green). Solid (dashed) lines are computed at $t=4$ orbits ($t=10$ orbits) of the closer companion. The spirals are located at $R\sim 180 R_{\rm g}$. Disc breaking is more successful in the case where the spirals are weaker as shown by the greater warp amplitude.}
    \label{fig:spirals_stop_breaking_profiles}
\end{figure}

Here we further investigate our simulation with $\kappa = 1.4 \times 10^{-4}$ and $\theta=40 \degr$, which we identify as unsuccessfully breaking. Figure~\ref{fig:spirals_stop_breaking_profiles} shows the evolution of the surface density and warp profiles throughout this simulation (purple curves). As the warp propagates from the inner region, it grows in amplitude with a peak at $R\sim130R_{\rm g}$,  until it reaches the region where the spiral arms are located. Upon meeting the spiral arms just before $R\sim 200R_{\rm g}$, the warp sharply decreases in amplitude and breaking is halted. This simulation thus suggests that the spiral arms can stabilise the disc against breaking when the disc is set to break near or beyond them. Such an effect is likely due to the local variations in the disc viscosity: spiral arms increase the local viscosity and a larger viscosity makes it harder to break the disc (e.g. Fig.~\ref{fig:break_notbreak}).  We note that this effect can be captured only using hydrodynamical simulations and is thus absent in the predictions of  \cite{2020MNRAS.496.3060G}.

\begin{figure}
    \centering
    \includegraphics[width=\columnwidth]{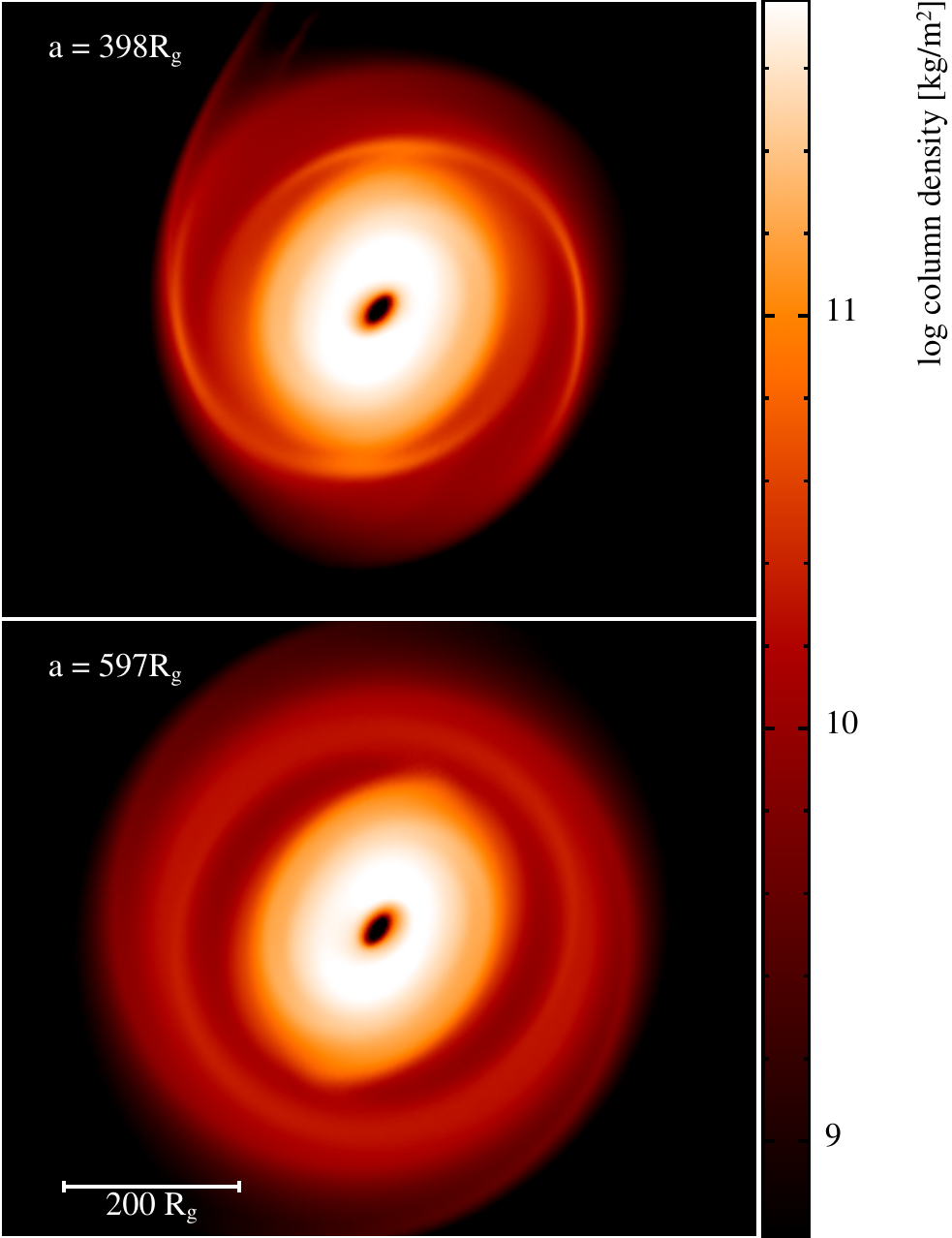}
    \caption{Breaking can be prevented by local hydrodynamical structures, like the spirals driven by the interaction with the binary companion. Upper and lower panel show a mass rendering of the simulations with $\theta=40\degr$ and $\kappa = 1.4 \times 10^{-4}, 4.3 \times 10^{-5}$, respectively, at the same timestep of the dotted lines in Fig.~\ref{fig:spirals_stop_breaking_profiles}. Increasing the binary separation reduces the strength of the spirals, making it easier to break the disc even though $\kappa$ decreases.}
    \label{fig:spirals_stop_breaking}
\end{figure}

To further investigate this behaviour, Fig.~\ref{fig:spirals_stop_breaking_profiles} also shows a second simulation (green curves) where the companion is placed at a larger radius $R_\star = 597 R_{\rm g}$ resulting in $\kappa = 4.3 \times 10^{-5}$. With the increased separation, the companion has a weaker effect on the disc, the spiral arms are weaker, and one would anticipate that it is harder to break the disc. However, Fig.~\ref{fig:spirals_stop_breaking_profiles} shows that the disc breaking is actually more successful ($\psi\sim 1.5$) in the case with the larger separation and weaker spiral arms. Figure~\ref{fig:spirals_stop_breaking} displays the density rendering of these comparison simulations at the same time as the final time-step shown on Fig.~\ref{fig:spirals_stop_breaking_profiles}. In the case with the strong spiral arms ($R_\star=398R_{\rm g}$) we find only unsuccessful breaking, whilst in the case with weak spiral arms ($R_\star = 597R_{\rm g}$) there is a break forming in the disc --- indeed roughly at the location of the spiral arms. 

This point illustrates that, while the parameter $\kappa$ captures the qualitative occurrence of the breaking, a more complex dynamics is present and can only be captured with  detailed  simulations. 
 
\subsection{Disc-black hole alignment}
\label{section:disc_BH_alignment}
\begin{figure}
    \centering
    \includegraphics[width=\columnwidth]{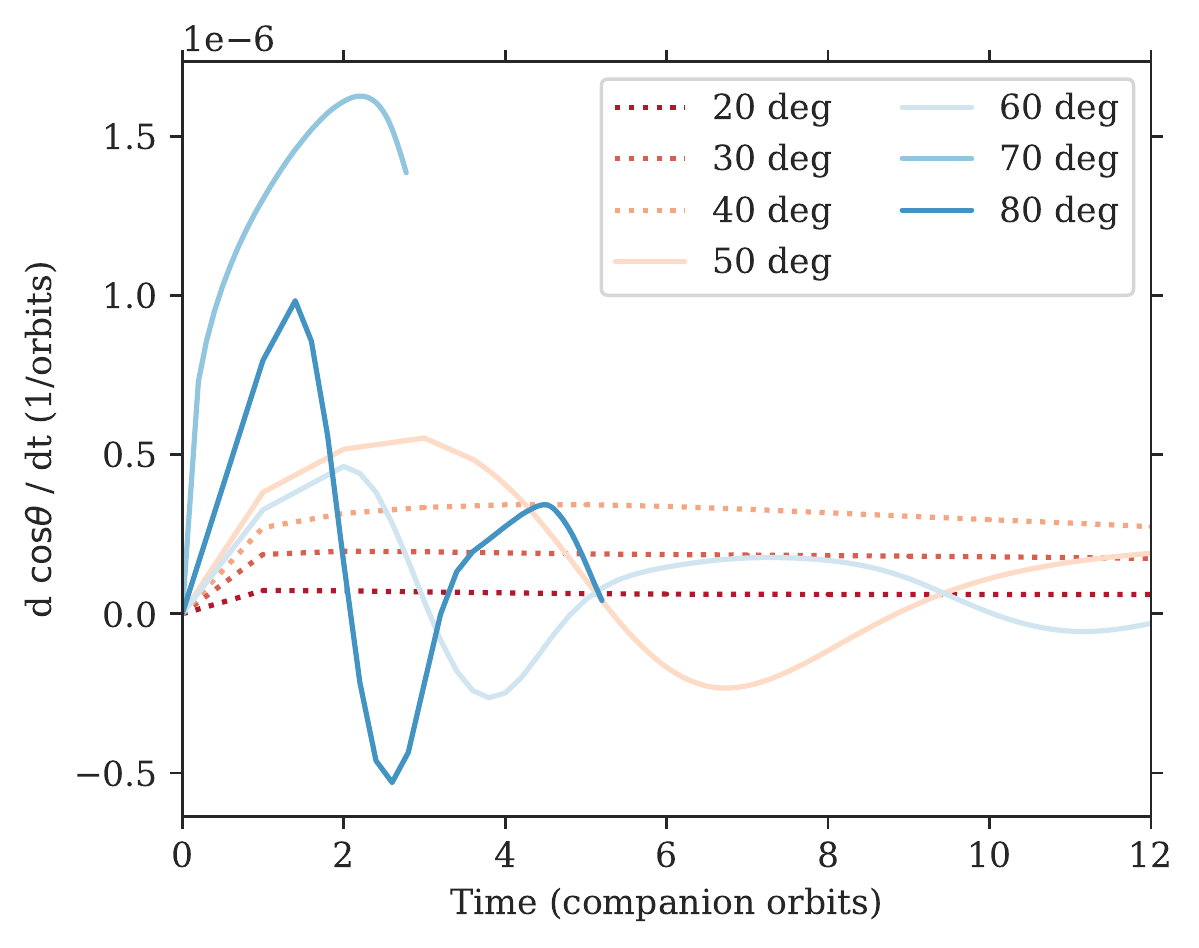}
    \includegraphics[width=\columnwidth]{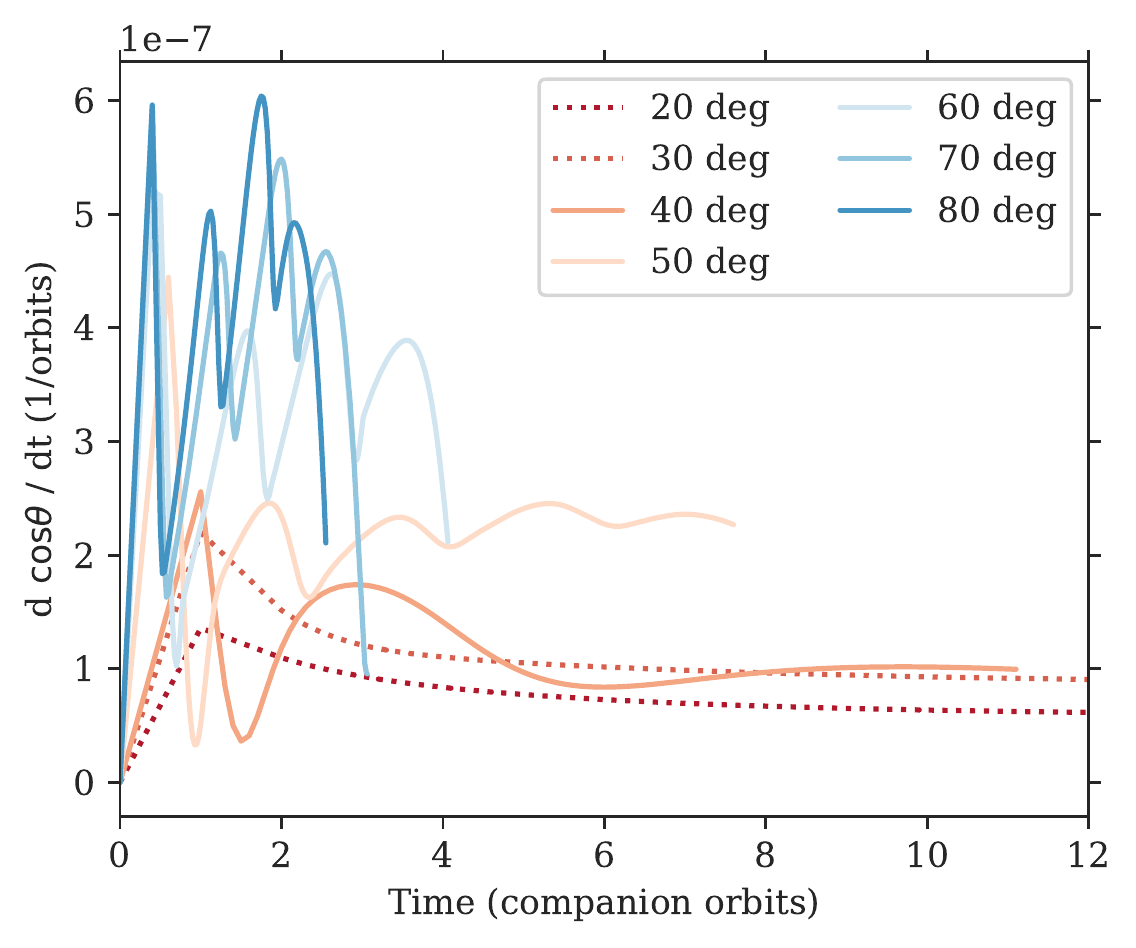}
    \caption{Predicted BH-disc alignment for simulations with one break (upper panel, $\kappa=1.4 \times 10^{-4}$) and multiple breaks (lower panel, $\kappa=2.6 \times 10^{-1}$). In each panel discs that have not broken are shown with a dotted line. The oscillatory profiles reflect the precession of the broken disc segments. Here we only show the prograde cases for clarity but note that the retrograde cases are qualitatively similar.} 
    \label{fig:dcdt}
\end{figure}

We now calculate $d \cos \theta/dt$ (cf. Sec.~\ref{backr} and Appendix~\ref{section:calculating_dcdt}) to evaluate how disc breaking affects alignment between the disc and BH. In Fig.~\ref{fig:dcdt} we show two representative sub-sets of our simulations with all discs prograde and $\kappa$ equal to either $1.4 \times 10^{-4}$ or $2.6 \times 10^{-1}$. While both sets contain simulations that have both warping and successful breaking, the discs in the low $\kappa$ set only include discs with a single break while the high $\kappa$ only presents multiple breaks.

When the disc remains warped with no signs of breaking (dotted lines, Fig.~\ref{fig:dcdt}), $d \cos \theta/dt$ is an increasing function that appears to asymptote to a value that is in fair agreement from the prediction from \citet{2020MNRAS.496.3060G}. As soon as a disc shows signs of successful breaking (solid lines), our measure for $d \cos \theta/dt$ develops strong oscillations on a timescale set by the precession of the inner broken disc, with $d \cos \theta/dt$ at its lowest when the inner disc(s) most strongly oppose the outer disc.

The amplitude of the oscillations depends on whether the disc has broken into two smaller discs (upper panel, low $\kappa$) or multiple rings (lower panel, high $\kappa$). We can understand this by considering Eq.~(\ref{equation:dJdt}), where we integrate from the inner edge to the outer edge of the disc. When the inner disc breaks and precesses, it can develop opposing angular momentum to the outer disc. Thus when we evaluate Eq.~(\ref{equation:dJdt}), it is possible that the contribution of the inner disc cancels out part of the contribution of the outer disc in the integral --- particularly when we take into account the $R^{-3}$ dependence of the Lense-Thirring torque. 

In the case that there is a single break, the inner disc has a relatively large radial extent and thus holds an appreciable fraction of the total disc angular momentum. Our estimation of $d \cos \theta/dt$ for these discs is mostly governed by the orientation of the inner disc and has a large amplitude.
If instead there are multiple rings, each of these is radially narrower and so hold a smaller fraction of the total disc angular momentum compared to the single-break case. Additionally, they  precess with a rate determined by their radius and so have a range of orientations. Configurations where they oppose each other are also possible. This leads to an oscillatory modulation on the $d \cos \theta/dt$ profile, with the oscillations reflecting the opposing angular momenta from the sum of the rings.

This behaviour is replicated across our parameter suite. For discs with one break we find that $d \cos \theta/dt$ is slowed and alignment between the disc and BH may be prevented. For discs with multiple rings, alignment between the outer disc and BH is hindered but not necessarily prevented.

\section{Discussion}
\label{section:discussion}

\subsection{Limitations of the 1D model}
Our comparison in Fig.~\ref{fig:break_notbreak} highlights the limitations of the semi-analytic model of \citet{2020MNRAS.496.3060G}. As expected, at large values of $\kappa$ the boundary assumptions of the 1D model mean we cannot compare directly. Additionally, we may expect non-axisymmetric effects to play a major role when the companion has such a strong influence. Here our simulations still recover combinations of stable and broken discs, with the broken discs at higher inclinations. Our results thus suggest that even when the companion has a strong influence, disc breaking can inhibit the alignment between the disc and the BH. At this extreme we do note that our assumption of a fixed potential may become problematic (but see Sec.~\ref{section:fixed_BH}).

We do not recover the predicted breaking for very thick ($H/R = 0.08$) or large viscosity discs ($\alpha=0.2$). In the former case, this is likely to be because the disc is thick enough that it has violated the assumption inherently made in the semi-analytic model that the disc is thin, even though the calculation of $\kappa$ depends on the aspect ratio. In the latter case this is likely due to the inner part of the disc accreting rapidly, preventing a ring from successfully breaking off.

\subsection{Numerical viscosity at low $\alpha$}
A potential cause for any discrepancy between the 1D semi-analytic model and the results of our simulations could be our modelled value of $\alpha$. While \citet{2010MNRAS.405.1212L} demonstrates that the $\alpha$ prescription in our simulations  is appropriate (down to smaller $\alpha$ than we have used here), this has not yet been confirmed in  presence of a break. Prior to 
breaking, the disc is continuous and resolved and so the viscosity treatment is robust. After the break has occurred, at the location of the break we have locally poor resolution which can naturally lead to higher local viscosity with our chosen viscosity implementation. In practice, this may mean that the low-$\alpha$ cases might appear to be more viscous than we expect, increasing the angle of criticality and moving points to larger $\kappa$ in Fig.~\ref{fig:break_notbreak}. We speculate this is the cause of the slight discrepancy observed for $\alpha=0.05$  and $\kappa=1.4 \times 10^{-4}$, but note that this does not appear to be an issue with any of our other sets.

\subsection{Assumption of a fixed potential}
\label{section:fixed_BH}
As introduced in Sec.~\ref{section:modelling_BH}, we use a fixed potential to model the primary rotating BH. Using a fixed potential in this manner is common; for example in other SPH codes \citep{2000MNRAS.315..570N}, in grid codes \citep{2020arXiv200812381D} and applications other than accretion discs \citep{2016MNRAS.455.2253B}. A consequence of this approach is that it is equivalent to assuming the primary BH is the centre of mass of the binary system. For the parameters we have chosen (see Table~\ref{tab:initial_conditions}), the centre of mass of our simulation is within $3.8R_{\rm g}$ of the primary BH. This is well within our numerical accretion radius of $15R_{\rm g}$ and so will not significantly impact the simulated disc evolution.

\subsection{Long term behaviour}
\label{section:longterm}
With (unfortunately) finite computational time at our disposal, we have chosen to run more simulations with different parameters for a shorter time rather than fewer simulations for longer times. This means that our simulations run for a timescale that is much shorter compared to both the viscous time of the disc or the binary inspiral time. However, for the vast majority of our simulations we do not require long time-scales to confirm if the disc is stable to breaking or not.

In the instance that the disc is unstable (i.e. in the region of the parameter space well beyond criticality), breaking occurs within a few orbits and we do not have to simulate further. Confirmation that breaking occurs on a few dynamical time-scales rather than the viscous time-scale can be inferred from Fig.~\ref{fig:dependence_on_viscosity}, where we see signatures of breaking for discs with high and low viscosities at the same number of companion orbits. If instead the disc is stable, a low amplitude $\psi(R)$ profile (e.g. Fig.~\ref{fig:tearing_example}) develops without local maxima. As the disc evolves, the $\psi(R)$ profile decreases in amplitude, moving the disc further towards stability \citep[e.g.][]{2018MNRAS.476.1519D}. With no changing external torques to influence the disc, this profile continues to decrease and it becomes increasingly stable against breaking. The difference in the evolution of the $\psi(R)$ profile for the stable/unstable cases is the primary feature we test for when we summarise our results in Fig.~\ref{fig:break_notbreak}.

The only discs in our suite that would benefit from longer simulation times are those that show unsuccessful breaking. We identify numerous cases of unsuccessful breaking where the disc is moving towards breaking, stalls and then subsequently stabilises. While we have made every effort to rule out that these discs are simply in the early stages of a successful break, the intriguing subtleties of this behaviour deserve further attention. 

\subsection{Isothermal equation of state}
The semi-analytic model of \citet{2020MNRAS.496.3060G} assumed a locally isothermal equation of state - that is, the temperature varies as a function of radius but does not allow for any disc heating. In order to compare with our 3D numerical simulations, we then assume the same equation of state which results in the disc thickness being kept constant at a given radius. While this is valid up until the point in our simulation where the disc breaks, it is not clear that this holds once the disc has broken. The broken disc geometry naturally leads to large relative velocities, and when these are of order of the Keplerian velocity these could easily lead to strong shocks. Such shocks may heat up the disc, altering the disc geometry and properties significantly.

Although assuming a particular equation of state is unlikely to be correct after the disc has broken, to date it is the most widely adopted approach \citep[e.g.][amongst others]{2000MNRAS.315..570N,2013MNRAS.433.2403O,2014MNRAS.441.1408T,2015MNRAS.448.1526N,2021MNRAS.507..983L,2020MNRAS.495.1148D}. Simulations that include radiative transfer will be able to account for heating due to shocks as well as any viscous heating or cooling that may occur in the broken disc. While this is not necessary for our comparison with the semi-analytic model and does not affect our assessment of whether the disc breaks or not, it should be taken into account when considering subsequent detailed evolution of broken discs.

\section{Conclusions}
\label{secconclusions}
In this work we have considered the structure of a misaligned accretion disc surrounding BHs in binary systems. With our suite of 143 SPH simulations we have shown that:
\begin{enumerate}[leftmargin=*]
    \item The `critical obliquity' first identified by \cite{2014MNRAS.441.1408T} and explored at length by \cite{2020MNRAS.496.3060G} does indeed correspond to disc breaking/tearing, where the disc separates into distinct segments.
    \item Our numerical simulations recover the qualitative predictions of the 1D semi-analytic model, with the mutual inclination that causes disc breaking decreasing with increasing $\kappa$ and decreasing $\alpha$.
    \item At the same time, 3D hydrodynamics allows us to unveil a richer phenomenology. Disc breaking hinders (and in some cases can prevents) alignment between the disc and the BH. The difference in this behaviour depends on the whether the disc breaks into two discs or multiple precessing rings.
    \item Hydrodynamic effects not taken into account by the semi-analytic model such as spiral arms are able to stabilise the disc against breaking.
\end{enumerate}

Our results have strong implications for inspiralling binary BHs. Most current models are based on the idea that the lighter binary member accretes more than its heavier companion (a.k.a. `differential accretion', e.g. \citealt{2015MNRAS.451.3941G,2020MNRAS.498..537S,2021MNRAS.501.2531S}), predicting that BH binaries should reach their merger phase with the primary's (secondary's) spins aligned (misaligned) with the orbital angular momentum of the binary. Once the critical obliquity is included in this picture, disc breaking implies that spin alignment is slowed if not completely prevented for a specific subset of systems. Although further modeling is  necessary to understand the full repercussions of our findings, this opens for the exciting prospect of exploiting future LISA measurements of precessing binary BHs to infer details on the dynamics of warped discs in gas-rich galaxies.

\section*{Acknowledgements}
The authors thank Nicola Giacobbo and the referee, Pavel Ivanov, for discussions and comments on the manuscript. 
R.N. acknowledges support from UKRI/EPSRC through a Stephen Hawking Fellowship (EP/T017287/1). 
E.R. acknowledges funding from the European Research Council (ERC) under the European Union’s Horizon 2020 research and innovation programme (grant agreement No 681601 and No 864965). 
D.G. is supported by European Union's H2020 ERC Starting Grant No. 945155--GWmining, Leverhulme Trust Grant No. RPG-2019-350, and Royal Society Grant No. RGS-R2-202004. 
G.R. acknowledges support from the Netherlands Organisation for Scientific Research (NWO, program number 016.Veni.192.233) and from an STFC Ernest Rutherford Fellowship (grant number ST/T003855/1).
This work was performed using the DiRAC Data Intensive service at Leicester, operated by the University of Leicester IT Services, which forms part of the STFC DiRAC HPC Facility. The equipment was funded by BEIS capital funding via STFC capital grants ST/K000373/1 and ST/R002363/1 and STFC DiRAC Operations grant ST/R001014/1. DiRAC is part of the National e-Infrastructure.
We also acknowledge computational resources from the University of Birmingham BlueBEAR cluster.
Figures were made using \textsc{splash} \citep{2007PASA...24..159P}.

\section*{Data Availability}
The data underlying this article will be shared on reasonable request to the corresponding author. The code Phantom used in this work
is publicly available at \url{https://github.com/danieljprice/phantom}.



\bibliographystyle{mnras_tex_edited}
\bibliography{bec} 



\appendix

\section{Spin-alignment implementation}
\label{section:calculating_dcdt}
Here we describe our method for calculating $d \cos \theta /dt$ as a function of time for each disc. Importantly, although Eq.~(\ref{equation:dJdt}) could in principle be calculated as a sum over particles, we instead calculate it over annuli as we describe here. We begin with a standard discretisation of the disc into $N$ concentric annuli, following \cite{2010MNRAS.405.1212L}. Here particles are binned into their respective annulus by their spherical radius because this accounts for any warped or misaligned disc structure. We assume azimuthal symmetry and average the properties of the particles in each annulus to recover the disc properties as a function of radius. While we note this is not strictly accurate as the spiral arms are not azimuthally symmetric, this best matches the 1D approach of \citet{2020MNRAS.496.3060G} which we are comparing to and the perturbations introduced by the spirals represent a small contribution to $d \cos \theta /dt$.

With this discretisation, one has $\Sigma_i$ and $\bm{\hat{l}_i}$ for each annulus $i$. We then write Eq.~(\ref{equation:dJdt}) in its similarly discretised form
\begin{align}
    \frac{d \bm{J}}{dt} = \frac{4 \pi G}{c^2} \sum_i^N \frac{ \bm{J} \times \bm{L}_i}{R_i^2}\,.
    \label{equation:summation}
\end{align}
where the angular momentum density is given by $\bm{L}_i = \Sigma_i \sqrt{GMR_i} \bm{\hat{l}_i}$ (note this is \emph{not} the same as $m \bm{r} \times \bm{v}$ calculated on each individual particle).
We then need to calculate the time derivative of the unit spin angular momentum, 
\begin{align}
    \frac{d \bm{\hat{J}}}{dt} = \frac{1}{J}\left(\frac{d\mathbfit{J}}{dt} - \bm{\hat{J}} \frac{dJ}{dt} \right)\,,
    \label{djhat}
\end{align}
which follows from $\bm{J} = J \bm{\hat{J}}$. The last term in Eq.~(\ref{djhat}) represents the change in the magnitude of the BH spin angular momentum. However, by definition, the Bardeen-Petterson torque ($d\bm J/dt$) is always perpendicular to $\bm J$. We thus use only the first term to calculate the alignment, with
\begin{align}
    \frac{\rm d \cos \theta}{\rm d t} =  \frac{1}{J}
    \frac{\rm d \mathbfit{J}}{\rm d t}  \cdot \hat{\mathbfit{L}}_*\,.
\end{align}
The last term is calculated from $L_* = M_{\star} \bm{R}_{\star} \times \bm{V}_{\star}$, where $V_{\star}$ is the binary's velocity relative to the primary. We complete the above procedure for each output of our simulation, to give $d \cos \theta /dt$ as a function of time for each disc, as shown in Fig.~\ref{fig:dcdt}.

\end{document}